\documentclass[showpacs,twocolumn,floats,eqsecnum,prb]{revtex4}
\usepackage{amssymb}

\usepackage{graphicx}

\begin{document}

\title{Odd spin-triplet superconductivity in a multilayered
superconductor-ferromagnet Josephson junction.}
\author{A.F. Volkov${}^{1,2}$}
\author{K. B. Efetov${}^{1}$}
\affiliation{$^{1}$Theoretische Physik III,\\
Ruhr-Universit\"{a}t Bochum, D-44780 Bochum, Germany\\
$^{2}$Institute for Radioengineering and Electronics of Russian Academy of\\
Sciences,11-7 Mokhovaya str., Moscow 125009, Russia}

\begin{abstract}
We study the dc Josephson effect in a diffusive multilayered SF'FF'S
structure, where S is a superconductor and F,F' are different ferromagnets.
We assume that the exchange energies in the F' and F layers are different ($%
h $ and $H$, respectively) and the middle F layer consists of two layers
with parallel or antiparallel magnetization vectors $M$. The $M$ vectors in
the left and right F' layers are generally not collinear to those in the F
layer. In the limit of a weak proximity effect we use a linearized Usadel
equation. Solving this equation, we calculate the Josephson critical current
for arbitrary temperatures, arbitrary thicknesses of the F' and F layers ($%
L_{h}$ and $L_{H}$) in the case of parallel and antiparallel $M$
orientations in the F layer. The part of the critical current $I_{cSR}$
formed by the short-range (SRC) singlet and $S=0$ triplet condensate
components decays on a short length $\xi _{H}=\sqrt{D/H}$, whereas the part $%
I_{cLR}$ due to the long-range triplet $|S|=1$ component (LRTC) decreases
with increasing $L_{H}$ on the length $\xi _{N}=\sqrt{D/\pi T}$. Our results
are in a qualitative agreement with the experiment \cite{Birge}.
\end{abstract}

\maketitle

\section{Introduction}

According to the Bardeen, Cooper and Schrieffer \cite{BCS} theory of
superconductivity in conventional metals and their alloys the
superconducting condensate consists of singlet Cooper pairs. These pairs can
be described by a wave function $f$ which is in the absence of the
condensate flow symmetric in the momentum space ($s$-wave singlet pairing).
In terms of the creation and annihilation operators $\psi ^{+},\psi ,$ this
function can be represented in the form of a thermodynamic average $%
f_{sng}(t-t^{\prime })\sim \langle \psi _{\uparrow }(t)\psi _{\downarrow
}(t^{\prime })-\psi _{\downarrow }(t^{\prime })\psi _{\uparrow }(t)\rangle .$
At equal times ($t=t^{\prime }$) this function determines the order
parameter $\Delta :$ $\Delta =\lambda \langle \psi _{\uparrow }(t)\psi
_{\downarrow }(t)\rangle =-\lambda \langle \psi _{\downarrow }(t)\psi
_{\uparrow }(t)\rangle $, where $\lambda $ is the coupling constant of an
attractive interaction$.$The spin-independent scattering by ordinary
impurities does not affect this type of superconductivity (Anderson-theorem) %
\cite{Schrieffer}.

In the last two decades other types of superconductivity have been
discovered. The most important example is high $T_{c}$ superconductivity
discovered by Bednorz and M\"{u}ller \cite{BednorzMuller} in cuprates.

As a result of intensive study of these superconductors, it was demonstrated
that, although the Cooper pairs in this case are also singlet, their wave
function essentially depends on the momentum $p$ and changes sign with
varying momentum direction in $CuO$ planes. In the simplest version, the
dependence of the order parameter $\Delta $ on $p$ has the form: $\Delta (%
\mathbf{p})=\Delta _{0}(\cos ^{2}(p_{x}a_{x})-\cos ^{2}(p_{y}a_{y}))$ ($d$%
-wave singlet pairing). Such a dependence $\Delta (\mathbf{p})$ allows one
to construct the so called $\pi -$Josephson junction consisting of single
crystals of high T$_{c}$ superconductors with an appropriate orientation of
these crystals with respect to each other \cite{Harlingen,Kirtley}. The
ground state of this junction corresponds to the phase difference equal to $%
\pi $.

Another type of superconductivity (triplet) has been discovered in strontium
ruthinate $Sr_{2}RuCu_{4}$ \cite{Maeno,Eremin} and in heavy fermion
intermetallic compounds \cite{Mineev}. In contrast to the singlet
superconductivity, the wave function of the Cooper pairs $f_{\uparrow
\uparrow }(\mathbf{p+p}^{\prime }\mathbf{,}t-t^{\prime })\sim \langle \psi
_{\uparrow }(\mathbf{p,}t)\psi _{\uparrow }(\mathbf{p}^{\prime }\mathbf{,}%
t^{\prime })\rangle $ is an odd function of momentum $\mathbf{p}$, so that
for equal times $t=t^{\prime }$ the function $f_{\uparrow \uparrow }(\mathbf{%
p,}0)$ and the order parameter $\Delta (\mathbf{p})$ do not equal to zero ($%
p $-wave triplet pairing). Only for some directions of the momentum $\mathbf{%
p} $ the order parameter turns to zero. This means that, in agreement with
the Pauli principle, the pair wave function changes sign under permutation
of spins and momenta. The momentum dependence of the condensate function
makes the singlet $d$-wave and triplet $p$-wave superconductivity sensitive
to scattering even by potential (not acting on spin) impurities.

An unusual mechanism of superfluidity was proposed by Berezinsky \cite{Berez}%
. Having in mind liquid $He^{3}$, he considered a retarded interaction
between atoms and assumed that the order parameter $\Delta (\omega )$ and
the wave functions $f_{\uparrow \uparrow }(\mathbf{p,}\omega )$ or $%
f_{\downarrow \downarrow }(\mathbf{p,}\omega )$ in the Matsubara
representation were even functions of momentum but odd function of the
Matsubara frequency $\omega $. However, experiments on superfluid $He^{3}$
revealed that the $p$-wave triplet type of the superfluidity is realized in $%
He^{3}$ rather than the one proposed by Berezinsky \cite{Leggett,Wolfle}.
Some possibilities to realize the exotic Berezinsky-type mechanism of
superconductivity in various systems in context of the pairing mechanism in
high $T_{c}$ superconductors were considered in Refs.\cite%
{Kirkpatrick,Coleman,Abrahams}.

This exotic type of superconductivity (or superfluidity) was regarded for
quite a long time as a hypothetical one. Only recently it has been realized %
\cite{BVE01} that the odd triplet superconductivity might exist in a simple
\ SF bilayer system consisting of a conventional\ $s$-wave singlet BCS-type
superconductor S and a ferromagnetic layer F with a nonhomogeneous
magnetization $M.$

To that moment it had already been well known that in an SF system with a
homogeneous magnetization $M$ the Cooper pairs penetrated the ferromagnet
over a short length $\xi _{F}=\sqrt{D/E_{ex}}$ (in the diffusive limit),
where $D=vl/3$ is the diffusion coefficient, $l=v\tau $ is the mean free
path, and $E_{ex}$ is the exchange energy in the ferromagnet (we set the
Planck constant $\hbar $ equal to $1$).\ Since the exchange energy usually
is much larger than the critical temperature of the superconductor $T_{c}$,
the length $\xi _{F}$ is much shorter than the length of the condensate
penetration into a normal metal in an SN bilayer: $\xi _{N}=\sqrt{D/\pi T}$ %
\cite{deGennes,GolubovRMP,BuzdinRMP,BVErmp}.

The Cooper pairs penetrating the ferromagnet with an uniform magnetization
consist of electrons with opposite spins. Their wave function is however the
sum of a singlet and triplet component with zero total spin projection on
the $z$-axis ($S=0$). The exchange field mixes these components and the
triplet component with $S=0$ is unavoidable in the ferromagnet. The sum of
these two components can be considered as a short-range component (SRC).

The part corresponding the triplet $S=0$ component has the form $%
f_{tr\uparrow \downarrow }(t-t^{\prime })\sim \langle \psi _{\uparrow
}(t)\psi _{\downarrow }(t^{\prime })+\psi _{\downarrow }(t^{\prime })\psi
_{\uparrow }(t)\rangle $. At equal times $t=t^{\prime }$ this function
equals zero in agreement with the Pauli principle. Therefore the function $%
f_{tr\uparrow \downarrow }(t-t^{\prime })$ is an odd function of the time
difference $(t-t^{\prime })$ or $\omega $ in the Matsubara representation.
The order parameter in the superconductor S is related only to the singlet
function $f_{sng}(\omega )$ which is an even function of $\omega .$ The
superconducting order parameter in F is zero if the coupling constant $%
\lambda _{F}$ in the Cooper channel equals zero.

The situation changes if the magnetization orientation in the vicinity of
the SF interface is not fixed. This case was analyzed in Ref. \cite{BVE01},
where an SF bilayer with a domain wall located at the SF interface was
considered. It was shown that in such a system not only the singlet and
triplet $S=0$ components but also the odd triplet component with $S=\pm 1$
arises in the ferromagnet. The latter component penetrates the
superconductor over a large distance that does not depend on the exchange
field and is of the order $\xi _{N}$ provided the spin-dependent scattering
is not too strong.

This odd triplet component can be considered as the long-range triplet
component (LRTC). As the LRTC is symmetric in the momentum space, the
scattering by potential impurities does not affect this component.

In subsequent theoretical papers various types of SF structures where the
LRTC may arise were studied (see review articles \cite%
{BVErmp,BuzdinRMP,EschrigRev} and references therein). In Ref. \cite{BVE01}
the creation of the LRTC is predicted in a diffusive SF structure with a
Bloch-type domain wall (DW). The width of the DW, $w,$ was assumed to be
larger than the mean free path $l:$ $l\ll $ $\{w,\xi _{F}\}.$

A more general case of the DW with a width, arbitrary with respect to the
mean free path, in a SF structure with an arbitrary impurity concentration
was studied in Ref. \cite{VE08}. The LRTC in diffusive SF structures with a
Neel-type DWs has been analyzed in Ref. \cite{Fominov05}. The case of a
half-metallic ferromagnet in SF or SFS structures was investigated in Refs. %
\cite{Eschrig03,AsanoBC,Zaikin08}. Braude and Nazarov \cite{Braude} studied
the LRTC in SF structures with a highly transparent SF interfaces so that
the amplitude of the condensate functions induced in the ferromagnet was not
small (strong proximity effect). Ballistic SF structures with a
nonhomogeneous magnetization, where the LRTC could be created, were studied
in Refs. \cite{Radovic,Beenakker,Valls,AsanoBC}. The papers \cite%
{VAE06,Champel08,Annett,SudboSp} were devoted to the study of the LRTC in
spiral ferromagnets attached to superconductors.

In several papers \cite{Eschrig03,AsanoBC,Sudbo,ZaikinBC,Belzig+Naz} the
LRTC was investigated in SF structures with the so-called spin-active
interfaces. In the approach used in these papers, the properties of the SF
interface are characterized by a scattering matrix with elements considered
as phenomenological parameters. In this approach one does not need knowing
the detailed structure of the SF interface and can proceed calculating
physical quantities using these parameters.\textbf{\ }We will see that even
in the framework of the quasiclassical theory one can obtain effective
boundary conditions for the LRTC provided the width of the DW $w$ attached
at the SF interface is thin enough (such an approach was used in Ref. \cite%
{VE08}). From the physical point of view the region with a narrow DW can be
regarded as a spin-active SF interface. If the width $w$ is comparable with
the Fermi wave length, one has to go beyond the quasiclassical theory and
derive the boundary conditions from the first principles (see \cite{Millis}
as well as \cite{EschrigBC09,Belzig+Naz} and references therein).

By now, several papers presenting a quite convincing experimental evidence
in favor of the existence of the LRTC have been published \cite%
{Petrashov,Klapwijk,Birge,Aarts10}. In Ref. \cite{Petrashov} the conductance of a
spiral ferromagnet ($Ho$) attached to two superconductors was measured. It
was concluded that the conductance variation below the superconducting
critical temperature $T_{c}$ is too large to be explained in terms of the
singlet component. Keizer et al. \cite{Klapwijk} observed the Josephson
effect in an SFS junction with a half-metallic ferromagnet $CrO_{2}$. The
thickness of the F layer was much larger (up to $\sim $1 mkm) than the
penetration depth of the short range condensate components. Moreover, in the
metal where free electrons with only one spin direction are allowed, no
pairs with opposite spins are possible. Therefore, only triplet $|S|=1$
component can survive in this ferromagnet \cite{Eschrig03}. However, there
is no controllable parameter in this system that would allow one to change
the amplitude of the LRTC. A similar long-range Josephson effect in a
SFS junction with $CrO_{2}$ as a ferromagnetic layer was observed by
Anwar et al. in a recent work \cite{Aarts10}.

Recently the dc long-range Josepshon effect has been observed in a more
complicated SFS structure with a controllable parameter \cite{Birge}. In the
experimental setup of this work, F was not a single ferromagnetic layer but
a multilayered structure of the NF'NFNF'N type, where N is a nonmagnetic
metal, F' is a weak ferromagnet ($PdNi$ or $CuNi$) and F is a strong
ferromagnet ($Co$). The middle F layer was in its turn a trilayer structure
consisting of two F layers with antiparallel orientation of magnetization $M$
and of a thin layer ($Ru$) that provides RKKY coupling between the F layers.

The authors of Ref. \cite{Birge} measured the Josephson critical current $%
I_{c}$ for different thicknesses $L$ of the F' and F layers (we denote the
thicknesses of the F' and F layers as $L_{h}$ and $L_{H}$ layers
respectively). It was demonstrated that in the absence of the F' layers ($%
L_{h}=0$) the critical current $I_{c}$ was negligible if the width of the F
layer $L_{H}$ essentially exceeded the small length $\xi _{H}=\sqrt{D/H}$,
where $H$ is the exchange energy in the F layer. This is what one expects
for the conventional superconductivity. However, adding the F' layers
resulted in an increase of the critical current $I_{c}$ by several orders.
The dependence of $I_{c}(L_{h})$ is non-monotonous: the critical current is
small at small and large $L_{h}$ reaching a maximum at $L_{h}\sim \xi _{h}$.

The authors of Ref.\cite{Birge} suggested an explanation of these results in
terms of the LRTC. Note that the mean free path $l$ in the structure studied
in Ref.\cite{Birge} is rather short (the diffusive limit in the F' layers
and an intermediate case in the F layer).

Theoretically the dc Josephson effect in multilayered SFS junctions with a
non-collinear magnetization orientation has been studied in several works.
In Refs. \cite{Radovic,Barash} the Josephson current in ballistic SFS
junctions was calculated. The diffusive SFF'S junctions were considered in
Refs. \cite{Ivanov,SudboJJ}. However the long-range Josephson effect in the
junctions with two F layers is not possible; the Josephson critical current $%
I_{c}$ is not exponentially small only if the total thickness of the
ferromagnetic layer, $L_{F}+L_{F^{\prime }}$, is comparable with the short
length $\xi _{F}$: $I_{c}\sim \exp (-(L_{F}+L_{F^{\prime }})/\xi _{F}). $

The diffusive Josephson junctions with three ferromagnetic layers and
non-collinear $M$ orientation, where the long-range Josephson coupling may
exist, have been analyzed in Refs. \cite{BVE03,BuzdinHouzet}. The authors of
Ref. \cite{BVE03} considered the F'SFSF'structure with different
magnetization $M$ orientations in the F and F' layers. In Ref. \cite%
{BuzdinHouzet} a somewhat different, but more suitable for experimental
realization, SF'FF'S structure with different $M$ directions in the F and F'
parts was analyzed. In both papers the exchange energy in the F and F' was
assumed to be equal.

The amplitude of the LRTC, $f_{1},$ and the Josephson critical current due
to this component $I_{cLR}$\ are calculated in both works. Although the
structures studied in Refs. \cite{BVE03,BuzdinHouzet} are different, the
results obtained are similar. The final formula for the critical current can
be written in both the cases as
\begin{equation}
I_{cLR}=F(L_{h})\sin \alpha _{l}\sin \alpha _{r}  \label{a1}
\end{equation}

In Eq. (\ref{a1}), $\alpha _{l,r}$ are angles between the $z$-axis and the
magnetization vectors in the left (right) F' layers, while the magnetization
$M$ in the F layer is assumed to be parallel to the $z$-axis. The function $%
F(L_{h})$ is a non-monotonic function with a maximum at $L_{h}\sim \xi _{h}$%
. This function vanishes at small and large thickness $L_{h}$ of the F'
layers (see Eq. (12) in Ref. \cite{BVE03} and Fig. 2 in Ref. \cite%
{BuzdinHouzet}).

Qualitatively, this prediction agrees with the observations in Ref. \cite%
{Birge}. However, experimental parameters presented in this publication are
well defined and this makes a more detailed comparison of the theoretical
predictions for LTRC with the experimental results quite interesting.

In this paper we analyze an SF'FF'S structure which, being formally similar
to that considered by Houzet and Buzdin \cite{BuzdinHouzet}, is in many
respects different and closer to the structure studied experimentally.

First, unlike Ref.\cite{BuzdinHouzet}, we assume that the exchange energies
in the F' and F layers ($h$ an $H$ respectively) are different.

Secondly, we assume that the SF' interface is not perfect and the proximity
effect is weak. This assumption allows us to linearize the Usadel equation
and to calculate the critical Josephson current $I_{c}$ at any temperatures $%
T$ (in Ref. \cite{BuzdinHouzet} only the case of temperatures close to $%
T_{c} $ was considered).

Thirdly, we also analyze the case when the F layer consists of two domains
with parallel and antiparallel orientations of magnetization.

At last, we derive a formula for the current $I_{c}$ for arbitrary
thicknesses, $L_{h,H}$, of the F' and F layers (in Ref. \cite{BuzdinHouzet}
a formula for $I_{c}$ is presented only in the limit $L_{h}/\xi _{h}\ll 1$).

The paper is organized as follows. In Section \ref{basic} we formulate the
problem and write down necessary equations. In Section \ref{thin} we analyze
the case of thin F' layers ($L_{h}/\xi _{h}\ll 1$). We calculate the
amplitudes of the short range singlet and triplet $S=0$ components as well
as the LRTC. In Section \ref{arbitrary} the case of arbitrary lengths $%
L_{h,H}$ will be considered under assumption that the angles $\alpha _{l,r}$
are small. Using the formulas obtained for the amplitudes of different
components, we calculate in Section \ref{current} the critical current $%
I_{c} $ in the limiting cases: a) for parallel and antiparallel $M$
orientations in the F layer and for arbitrary angles $\alpha _{l,r}$
assuming that the thickness $L_{h}$ is small, b) for arbitrary $L_{h,H}$
under assumption of small $\alpha _{l,r}$. The results obtained are
discussed in Section \ref{discussion}.

\section{Model and Basic Equations\label{basic}}

We consider a multi-layer S/F Josephson junction shown schematically in
Fig.1. It consists of two superconductors, S, and three ferromagnetic layers
F, F'$_{\text{r,l}}$. The middle F layer may consist of two domains or
layers with parallel or antiparallel orientations of the magnetization $M$.
A similar Josephson junction has been studied experimentally in a recent
work \cite{Birge}.

The presence of normal N layers in the experimental S/NF$_{\text{l}}$NFNF$_{%
\text{r}}$N/S structures cannot change qualitatively the results for the S/F'%
$_{\text{l}}$FF'$_{\text{r}}/$S structure obtained here because the
scattering in the N layers does not depend on spin (if a weak spin-orbit
scattering can be neglected). Therefore, all the superconducting components,
singlet and triplet, decay in the N layers in a similar way over a large
distance of the order $\xi _{N}$. The exchange fields acting on electron
spins are $h$ in the F$^{\prime }$ layers and $H$ in the middle F layer. The
magnetization vector $\mathbf{M}$ in F is supposed to align along the z-axis
and it has the components $M(0,\sin \alpha _{l,r},\cos \alpha _{l,r})$ in
the F'$_{\text{l,r}}$ layers. The magnetization in the F layer is oriented
along the $z$-axis but may have parallel or antiparallel orientations in the
regions ($-L_{H}<x<0$) and ($0<x<L_{H}$).

\begin{figure}[tbp]
\begin{center}
\includegraphics[width=7cm, height=6cm]{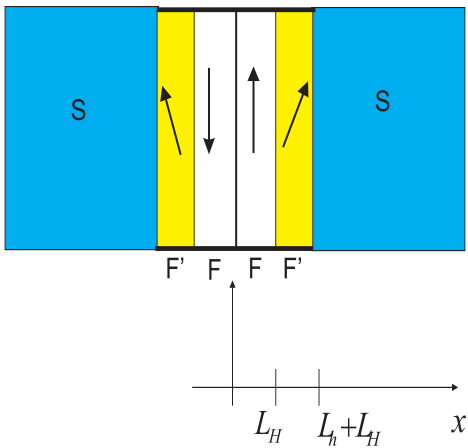}
\end{center}
\caption{(color online) Josephson structure under consideration. The F'(F)
layers are weak (strong) ferromagnets. The middle F layer consists of two
layers with parallel or antiparallel (shown in figure) magnetization
orientation. The arrows show the directions of the magnetization in F' and F
layers.}
\end{figure}

For explicit calculations we use the quasiclassical\ Green's function
technique, which is the most efficient tool for studying SF structures (see
reviews \cite{LOrev,KopninRev,Rammer,BelzigRev,GolubovRMP,BuzdinRMP,BVErmp}%
), and assume that all the ferromagnetic layers are in the diffusive regime,
so that the Usadel equation can be used. The amplitude of the condensate
wave function in the ferromagnetic layers is assumed to be small (weak
proximity effect) and therefore the Usadel equation can be linearized. The
smallness of the condensate wave function is either due to a mismatch of the
Fermi velocities in S and F or due to the presence of a tunnel barrier at
the S/F interfaces.

The anomalous (Gor'kov) quasiclassical Green's function in the considered
case of a spin-dependent interaction is a $4\times 4$ matrix $\check{f}$. We
are interested in the dc Josephson current $I_{c},$ i.e. in a
thermodynamical quantity. Therefore we can use the Matsubara representation
for the matrix $\check{f}$ \ and consider $\check{f}$ as a function of the
Matsubara frequency $\omega =\pi T(2n+1)$ and coordinate $x$ normal to
interfaces: $\check{f}=\check{f}(\omega ,x)$. The linearized Usadel equation
for $\check{f}$ has the form (see \cite{BVErmp}, Eq.(3.15))

\begin{eqnarray}
&&\partial ^{2}\check{f}/\partial x^{2}-\kappa _{\omega }^{2}\check{f}%
\;-i(\kappa _{F}^{2}/2)\cos \alpha (x)  \nonumber \\
&&\times \{\tan \alpha (x)\hat{\tau}_{3}\mathbf{\otimes }[\hat{\sigma}_{2},%
\check{f}]+[\hat{\sigma}_{3},\check{f}]_{+}\}=0  \label{Usadel}
\end{eqnarray}%
where $\kappa _{\omega }^{2}=2|\omega |/D$, $\kappa _{F}^{2}=2\mathit{sgn}%
\omega \cdot h/D$ in the F'$_{\text{l,r}}$ layers and $\kappa _{F}^{2}=2%
\mathit{sgn}\omega \cdot H/D$ in the F layer, the Pauli matrices $\hat{\tau}%
_{i}\mathbf{,}\hat{\sigma}_{i}$ operate in the particle-hole and spin space
respectively\textbf{.} The angle $\alpha $ is equal to $\alpha _{l,r}$ in
the F'$_{\text{l,r}}$ layers and to zero in the F layer in the case of the
parallel orientation of the magnetization $M$ in the domains. In the case of
the antiparallel orientation $\alpha (x)=\pi $ in the interval ($-L_{H}<x<0$%
) and $\alpha (x)=0$ in the interval ($0<x<L_{H}$). The diffusion
coefficient $D$ is assumed to be the same in all the ferromagnetic layers.

The matrix $\check{f}$ can be represented for the system under consideration
in a form of an expansion in the spin matrices $\hat{\sigma}_{i}$ as

\begin{equation}
\check{f}=\hat{f}_{0}\mathbf{\otimes }\hat{\sigma}_{0}+\hat{f}_{1}\mathbf{%
\otimes }\hat{\sigma}_{1}+\hat{f}_{3}\mathbf{\otimes }\hat{\sigma}_{3}
\label{fMatrix}
\end{equation}%
The matrices $\hat{\sigma}_{0}$ and $\hat{\sigma}_{1,3}$ are the unit matrix
and the $\hat{\sigma}_{x,z}$ Pauli matrices, respectively. The $\hat{f}%
_{0,1,3}$ matrices are matrices in the particle-hole space. The first term
is the short range triplet component with the zero projection of the total
spin on the z-axis, the second term is the LRTC with the non-zero projection
of the total spin, and the third term is the singlet component of the
condensate Green's function (see \cite{BVE03,BVErmp}).

Eq. (\ref{Usadel}) should be complemented by boundary conditions. We
consider the simplest model of the S/F heterostructures assuming that the
interfaces have no effect on spins (spin-passive interface). These boundary
conditions have the form \cite{ZaitsevBC,KL}

\begin{equation}
\partial \check{f}/\partial x|_{x=\pm L}=\pm \gamma _{B}\check{f}%
_{S}|_{x=\pm L},  \label{BC}
\end{equation}%
where $\gamma _{B}=1/(R_{B}\sigma ),R_{B}$ is the S/F interface resistance
per unit area, $\sigma $ is the conductivity of the ferromagnet. The matrix $%
\check{f}_{S}$ is the Gor'kov's quasiclassical Green's function in the left
and right superconductors. It has the form

\begin{equation}
\check{f}_{S}|_{x=\pm L}=f_{S}\hat{\sigma}_{3}\mathbf{\otimes }(\hat{\tau}%
_{2}\cos \varphi \pm \hat{\tau}_{1}\sin \varphi ),  \label{fS}
\end{equation}%
where $f_{S}=\Delta /\sqrt{\omega ^{2}+\Delta ^{2}},$ $\pm \varphi $ is the
phase in the right (left) superconductor, so that the phase difference is $%
2\varphi $.

If there is a spin-dependent interaction in a thin layer at the interface
(exchange field, spin-dependent scattering etc), the boundary condition
acquires a more complicated form. In particular, the coefficient $\gamma
_{B} $ becomes a matrix with matrix elements containing very often unknown
phenomenological parameters. Such interfaces are called spin-active
interfaces. In many papers the LRTC is studied in SF systems with
spin-active interfaces \cite{Eschrig03,AsanoBC,Sudbo,Zaikin08,Belzig+Naz}.

The F/F'$_{\text{l,r}}$ interfaces are assumed to be ideal and therefore the
function $\check{f}(x)$ and its derivative $\partial \check{f}/\partial x$
must be continuous at these interfaces. Solving the linear equation (\ref%
{Usadel}) with the boundary conditions (\ref{BC}) one can calculate the dc
Josephson current using the formula \cite{BuzdinRMP,BVErmp}

\begin{equation}
j_{J}=(\sigma /8)2\pi T\sum_{\omega \geq 0}Tr\{\hat{\sigma}_{0}\mathbf{%
\otimes }\hat{\tau}_{3}\check{f}\partial \check{f}/\partial x\},
\label{JosCurrent}
\end{equation}

This problem can be solved in a general case of an arbitrary thicknesses of
the F and F' layers ($L_{H}$ and $L_{h}$) and angles $\alpha _{l,r}$.

However, the general results are too cumbersome. In order to present
analytical formulas in a more or less compact form, we consider two limiting
cases: a) thin F'$_{\text{l,r}}$ layers ($L_{h}\ll \xi _{h},\xi _{N}$) and
arbitrary angles $\alpha _{r,l}$ , b) arbitrary thicknesses $L_{H},L_{h}$,
but small angles $\alpha _{r,l}$ ($\alpha \ll 1$). In the next section we
consider the case a).\bigskip

\section{Thin F$^{\prime }$ layers\label{thin}}

In this section we assume that the F'$_{\text{l,r}}$ layers (or $h$-layers)
are very thin so that the inequality $|\kappa _{h}|L_{h}\ll 1$ is satisfied,
where $\kappa _{h}^{2}=2\mathit{sgn}\omega (h/D)$ (usually $\kappa _{h}\gg
\kappa _{N}=\sqrt{\pi T/D}$ and therefore the condition $\kappa _{N}L_{h}\ll
1$ is also fulfilled). In the zero order approximation in the parameter $%
\kappa _{h}L_{h},$ the exchange field in the entire ferromagnetic region
except the thin F'$_{\text{l,r}}$ layers is homogeneous and equal to $H$.
Thus, only the $\hat{f}_{0,3}$ components in the expansion (\ref{fMatrix})
that describe the short-range components (SRC) differ from zero.

The matrix $\check{f}$ satisfies the equation

\begin{equation}
\partial ^{2}\check{f}/\partial x^{2}-\kappa _{\omega }^{2}\check{f}%
\;-i\kappa _{H}^{2}\cos \alpha (x)\hat{\sigma}_{3}\mathbf{\otimes }\check{f}%
=0,  \label{Usadel1}
\end{equation}%
where $\kappa _{H}^{2}=2\mathit{sgn}\omega (H/D)$.

The angle $\alpha (x)=0$ in the case of parallel orientation of the vector $%
\mathbf{M,}$ and $\alpha (x)=0$ at $x>0$, $\alpha (x)=\pi $ at $x<0$ in the
case of antiparallel orientation. We rewrite Eq. (\ref{Usadel1}) for the
diagonal in spin space components $\hat{f}_{\pm }=\hat{f}_{0}\pm \hat{f}_{3}$
as

\begin{eqnarray}
\partial ^{2}\hat{f}_{\pm }/\partial x^{2}-\kappa _{\pm }^{2}\hat{f}_{\pm }
&=&0,\text{ \ }x>0,  \label{Usadel2a} \\
\partial ^{2}\hat{f}_{\pm }/\partial x^{2}-\hat{\bar{\kappa}}_{\pm }^{2}\hat{%
f}_{\pm } &=&0,\text{ \ }x<0,  \label{Usadel2b}
\end{eqnarray}%
where $\kappa _{\pm }^{2}=\kappa _{\omega }^{2}\pm i\kappa _{H}^{2},$ $\hat{%
\bar{\kappa}}_{\pm }^{2}=\kappa _{\pm }^{2}$ in the case of the parallel $M$
orientation in both domains ($x>0$ and $x<0$) and $\hat{\bar{\kappa}}%
_{H}^{2}=\kappa _{\omega }^{2}\mp i\kappa _{H}^{2}$ if the magnetization
vector at $x<0$ changes sign with respect to its direction at $x>0$. The
boundary conditions for matrices $\hat{f}_{\pm }$ follow from Eq. (\ref{BC})

\begin{eqnarray}
\partial \hat{f}_{\pm }/\partial x& =&\pm \gamma _{B}f_{S}(\hat{\tau}%
_{2}\cos \varphi +\hat{\tau}_{1}\sin \varphi ),\text{ }  \nonumber \\
\text{\ }x &=&L_{H},  \label{BC1a} \\
\partial \hat{f}_{\pm }/\partial x &=&\mp \gamma _{B}f_{S}(\hat{\tau}%
_{2}\cos \varphi -\hat{\tau}_{1}\sin \varphi ),\text{ }  \nonumber \\
\text{\ }x &=&-L_{H};  \label{BC1b}
\end{eqnarray}

The solution of Eqs.(\ref{Usadel2a}-\ref{Usadel2b}) can be sought in the form

\begin{eqnarray}
\hat{f}_{\pm }(x) &=&\hat{A}_{\pm }\cosh (\kappa _{\pm }x)\cos \varphi +\hat{%
B}_{\pm }\sinh (\kappa _{\pm }x)\sin \varphi ,\text{ }  \nonumber \\
\text{\ }x &>&0,  \label{fpm>0} \\
\hat{f}_{\pm }(x) &=&\hat{\bar{A}}_{\pm }\cosh (\bar{\kappa}_{\pm }x)\cos
\varphi +\hat{\bar{B}}_{\pm }\sinh (\bar{\kappa}_{\pm }x)\sin \varphi ,\text{
}  \nonumber \\
\text{\ }x &<&0,  \label{fpm<0}
\end{eqnarray}

The relations between coefficients $\hat{A},\hat{\bar{A}}$ and $\hat{B},\hat{%
\bar{B}}$ should be found from the continuity of the matrices $\hat{f}_{\pm
}(x)$ and their derivatives $\partial \hat{f}_{\pm }/\partial x$ at $x=0$.
This gives: $\hat{A}=\hat{\bar{A}},$ $\kappa _{H}\hat{B}=\bar{\kappa}_{H}%
\hat{\bar{B}}.$ Using the boundary conditions (\ref{BC1a},\ref{BC1b}), we
find the coefficients $\hat{A}_{\pm },\hat{B}_{\pm }$

\begin{eqnarray}
\hat{A}_{\pm } &=&\pm \frac{\gamma _{B}}{\mathcal{D}_{\pm }}[\hat{f}%
_{S}(L_{H})\cosh \bar{\theta}_{\pm }+\hat{f}_{S}(-L_{H})\cosh \theta _{\pm
}],  \nonumber \\
\hat{B}_{\pm } &=&\pm \frac{\gamma _{B}}{\mathcal{D}_{\pm }}[\hat{f}%
_{S}(L_{H})\frac{\bar{\kappa}_{\pm }}{\kappa _{\pm }}\sinh \bar{\theta}_{\pm
}-\hat{f}_{S}(-L_{H})\sinh \theta _{\pm }],  \nonumber \\
&&  \label{Bpm}
\end{eqnarray}%
where $\mathcal{D}_{\pm }=\kappa _{\pm }\sinh \theta _{\pm }\cosh \bar{\theta%
}_{\pm }+\bar{\kappa}_{\pm }\sinh \bar{\theta}_{\pm }\cosh \theta _{\pm },$ $%
\theta _{\pm }=\kappa _{\pm }L_{H}$ and $\bar{\theta}_{\pm }=\bar{\kappa}%
_{\pm }L_{H}.$

In the case of parallel (P) and antiparallel (AP) orientations of the
magnetization in the F layer we obtain the function $\mathcal{D}_{\pm }$

\begin{eqnarray}
\mathcal{D}_{\pm P} &=&2\kappa _{\pm }\sinh \theta _{\pm }\cosh \theta _{\pm
}\text{ },  \nonumber \\
\text{ }\mathcal{D}_{+AP} &=&\mathcal{D}_{-AP}=2Re(\kappa _{+}\sinh \theta
_{+}\cosh \theta _{-}).  \nonumber \\
&&  \label{Den}
\end{eqnarray}%
One can see from Eqs. (\ref{fpm>0}-\ref{Bpm}) that the SRC decays
exponentially away from the SF interfaces over the short length $\xi _{H}.$
Indeed, at $|L_{H}-x|\gg \xi _{H}$ we obtain from Eqs. (\ref{fpm>0}-\ref{Bpm}%
) that $\hat{f}_{\pm }(x)\sim (\gamma _{B}/\kappa _{\pm })f_{S}(\hat{\tau}%
_{2}\cos \varphi +\hat{\tau}_{1}\sin \varphi )\exp (-(L_{H}-x)/\xi _{H}).$

Let us now find the LRTC. First, we obtain the effective boundary conditions
for this component. Assuming that $L_{h}\ll \xi _{h}$, we can integrate Eq.(%
\ref{Usadel}) over the thickness of the F'-layers and come to effective
boundary conditions for the triplet component

\begin{equation}
\partial \hat{F}_{1}/\partial x|_{x=\pm L_{H}}=\pm \gamma _{1}\hat{f}%
_{3}(\pm L_{H})\sin \alpha _{r,l},\text{ }  \label{efBC}
\end{equation}%
where $\gamma _{1}\equiv \kappa _{h}^{2}L_{h}$.

We have introduced in Eq. (\ref{efBC}) a matrix $\hat{F}_{1}=\hat{\tau}_{3}%
\mathbf{\otimes }\hat{f}_{1}$ describing the LRTC. This matrix $\hat{F}_{1}$
satisfies an equation that directly follows from Eq.(\ref{Usadel})

\begin{equation}
\partial ^{2}\hat{F}_{1}/\partial x^{2}-\kappa _{\omega }^{2}\hat{F}_{1}=0.
\label{Eqf1}
\end{equation}%
The solution for the matrix $\hat{F}_{1}$ can be written as

\begin{equation}
\hat{F}_{1}=\hat{A}_{1}\cosh (\kappa _{\omega }x)+\hat{B}_{1}\sinh (\kappa
_{\omega }x)  \label{f1}
\end{equation}%
\bigskip From the effective boundary conditions (\ref{efBC}) we find

\begin{eqnarray}
\hat{A}_{1} &=&\frac{\gamma _{1}}{2\kappa _{\omega }sinh\theta _{\omega }}[%
\hat{f}_{3}(L_{H})\sin \alpha _{r}+\hat{f}_{3}(-L_{H})\sin \alpha _{l}],
\nonumber \\
\hat{B}_{1} &=&\frac{\gamma _{1}}{2\kappa _{\omega }cosh\theta _{\omega }}[%
\hat{f}_{3}(L_{H})\sin \alpha _{r}-\hat{f}_{3}(-L_{H})\sin \alpha _{l}].
\nonumber \\
&&  \label{A1B1}
\end{eqnarray}%
where $\gamma _{1}=2\mathit{sgn}\omega (h/D)L_{h}$, i.e. the matrix $\hat{F}%
_{1}$ is an odd function of the Matsubara frequency.

The solution for $\hat{F}_{1}$, Eq. (\ref{f1}), demonstrates that the LRTC
described by the function $\hat{F}_{1}$ decays slowly at a large distance of
the order $\kappa _{\omega }^{-1}\sim \xi _{N}$. The matrix $\hat{f}_{3}$ in
Eq. (\ref{A1B1}) is expressed through $\hat{f}_{\pm }:$ $\hat{f}_{3}=(\hat{f}%
_{+}-\hat{f}_{-})/2$, where the matrices $\hat{f}_{\pm }(\pm L_{H})$ are
given by Eqs. (\ref{fpm>0}-\ref{Bpm}).

As it should be, the function $\hat{F}_{1}(x)$ turns to zero in the absence
the exchange field or in the case of collinear magnetization because $\gamma
_{1}=\kappa _{h}^{2}L_{h}\sim h$ and $\sin \alpha _{r,l}=0$ in the case of
collinear $M$ orientations.

Eqs. (\ref{fpm>0}-\ref{Bpm},\ref{f1}-\ref{A1B1}) determine all the
condensate Green's functions. Using these functions, we calculate the
Josephson current in Section \ref{current}.

\section{Arbitrary thicknesses of ferromagnetic layers at weak
non-collinearity\label{arbitrary}}

Consider now a more interesting case of an arbitrary thicknesses of the
ferromagnetic layers F$^{\prime }$, F (or $h$, $H$ - layers). We restrict
ourselves with the case of the parallel $M$ orientations in the F layer
because there is no qualitative difference between the behaviour of the LRTC
in the P and AP magnetic configurations. For simplicity we assume that the
angle $\alpha $ is small, $\alpha \ll 1$. In this case the amplitude of the
LRTC is proportional to the small parameter $\alpha .$ In the zero order
approximation only the singlet component, $\hat{f}_{3}$ , and the short
range triplet component, $\hat{f}_{0}$, with zero projection of the total
spin of Cooper pairs on the z-axis are not zero. Indeed, we will look for a
solution of Eq. (\ref{Usadel}) in the form $\hat{f}_{3}(x)\sim $ $\hat{f}%
_{0}(x)\sim $ $\hat{f}_{1}(x)\sim $ $\{\cosh (\kappa x),\sinh (\kappa x)\}$,
where $\kappa $ is the eigenvalue.

In the ferromagnetic layers we obtain the following equations for the
eigenvectors

\begin{eqnarray}
\hat{f}_{0}(\kappa ^{2}-\kappa _{\omega }^{2})-\hat{f}_{3}i\kappa
_{F}^{2}\cos \alpha &=&0  \label{Eigen1} \\
\hat{f}_{3}(\kappa ^{2}-\kappa _{\omega }^{2})-\hat{f}_{0}i\kappa
_{F}^{2}\cos \alpha -\hat{F}_{1}\kappa _{F}^{2}\sin \alpha &=&0
\label{Eigen2} \\
\hat{F}_{1}(\kappa ^{2}-\kappa _{\omega }^{2})+\hat{f}_{3}\kappa
_{F}^{2}\sin \alpha &=&0  \label{Eigen3}
\end{eqnarray}%
where the matrix $\hat{F}_{1}$ introduced in Eq.(\ref{efBC}) describes the
LRTC. This set of equations has three eigenvalues

\begin{equation}
\kappa _{1,2}^{2}\equiv \kappa _{F\pm }^{2}=\kappa _{\omega }^{2}\pm i\kappa
_{F}^{2},\text{ }\kappa _{3}^{2}=\kappa _{\omega }^{2}  \label{Eigenvalues}
\end{equation}

Two of them, $\kappa _{F\pm }$ , describe a sharp decay of the density of
Cooper pairs in the ferromagnet (in the case $\{H,h\}\gg T,\Delta $) and the
latter one, $\kappa _{\omega }$ ($\kappa _{\omega }=1/\xi _{LR}$), is an
inverse characteristic length of decay of the LRTC in the ferromagnet. By
order of magnitude it is equal to $\kappa _{\omega }^{2}\approx \pi T/D,$
which shows that the length $\xi _{LR}$ is rather large and does not depend
on the exchange energies $h,H$. Spin-orbit interaction or a spin-dependent
impurity scattering make this length shorter \cite%
{BVErmp,BuzdinRMP,Demler,Blanter2,BVE07,IvanovFomin,Skvortsov}

\begin{equation}
\kappa _{3}^{2}=\kappa _{\omega }^{2}+\kappa _{m}^{2}  \label{MagScat}
\end{equation}%
where $\kappa _{m}^{-2}\approx \min \{D\tau _{m},D\tau _{sp-orb}\}$, $\tau
_{m}$ and $\tau _{sp-orb}$ are characteristic times related to the
spin-dependent impurity scattering or spin-orbit interaction. The lengths $%
\kappa _{F\pm }^{-1}$ also depend on $\kappa _{m}^{2}$ and can be found by
shifting $\kappa _{F\pm }^{2}\Rightarrow \kappa _{F\pm }^{2}+\kappa
_{mF}^{2} $.

It is seen from Eqs. (\ref{Eigen1}-\ref{Eigen3}) that the LRTC arises only
at non-zero $\alpha $ when $\hat{F}_{1}\neq 0$. In the zero-order
approximation ($\alpha =0$) we should find the matrices $\hat{f}_{0,3}$ in
each ferromagnetic layer. As follows from Eqs. (\ref{Eigen1}-\ref{Eigen2}),
at $\alpha =0$ only the eigenvectors $\hat{f}_{0\pm }=\pm \hat{f}_{3\pm }$
corresponding to the eigenvalues $\kappa _{\pm }$ can be finite.

The solution for $\hat{f}_{0,3}$ satisfying the boundary conditions (\ref{BC}%
) can be written as

\begin{eqnarray}
\hat{f}_{3}(x) &=&\sum_{j}\{\hat{\tau}_{2}\mathcal{A}_{Hi}\cosh (\kappa
_{Hj}x)\cos \varphi  \nonumber \\
&&+\hat{\tau}_{1}\mathcal{B}_{Hj}\sinh (\kappa _{Hj}x)\sin \varphi \}
\label{f3H} \\
\hat{f}_{0}(x) &=&\sum_{j}(-1)^{j+1}\{\hat{\tau}_{2}\mathcal{A}_{Hj}\cosh
(\kappa _{Hj}x)\cos \varphi  \nonumber \\
&&+\hat{\tau}_{1}\mathcal{B}_{Hj}\sinh (\kappa _{Hj}x)\sin \varphi \}
\label{f0H}
\end{eqnarray}%
in the $H$-region (F layer) and

\begin{eqnarray}
&&\hat{f}_{3}(x)=\sum_{j}\{\hat{\tau}_{2}[C_{hj}^{(2)}\cosh (\kappa _{hj}%
\tilde{x})+S_{hj}^{(2)}\sinh (\kappa _{hj}\tilde{x})]  \nonumber \\
&&+\hat{\tau}_{1}[C_{hj}^{(1)}\cosh (\kappa _{hj}\tilde{x}%
)+S_{hj}^{(1)}\sinh (\kappa _{hj}\tilde{x})]\}  \label{f3h} \\
&&\hat{f}_{0}(x)=\sum_{j}(-1)^{j+1}\{\hat{\tau}_{2}[C_{hj}^{(2)}\cosh
(\kappa _{hj}\tilde{x})+S_{hj}^{(2)}\sinh (\kappa _{hj}\tilde{x})]  \nonumber
\\
&&+\hat{\tau}_{1}[C_{hj}^{(1)}\cosh (\kappa _{hj}\tilde{x}%
)+S_{Hj}^{(1)}\sinh (\kappa _{hj}\tilde{x})]\}  \label{f0h}
\end{eqnarray}%
in the $h$-region (F' layers), where $\tilde{x}=x-L$, $j=1,2$ so that $%
\kappa _{H1,2}=\kappa _{H\pm }$ etc.

The coefficients $\mathcal{A}_{Hi},\mathcal{B}_{Hi}$ and $%
C_{Hi}^{(1)},S_{Hi}^{(1)},C_{Hi}^{(2)},S_{Hi}^{(2)}$ are found from the
boundary conditions (\ref{BC}). We write down here the expressions for $%
C_{Hi},S_{Hi}$ (see Appendix)

\begin{eqnarray}
\mathcal{A}_{H\pm } &=&\lambda _{\pm }S_{h\pm }^{(2)}/R_{\pm }(cs;sc),
\label{CH} \\
\mathcal{B}_{H\pm } &=&\lambda _{\pm }S_{h\pm }^{(1)}/R_{\pm }(cc;ss)
\label{SH}
\end{eqnarray}%
where $\lambda _{\pm }=(\kappa _{h}/\kappa _{H})_{\pm }$ and $R_{\pm
}(cs;sc)=(\cosh \theta _{h}\sinh \theta _{H}+\lambda \sinh \theta _{h}\cosh
\theta _{H})_{\pm }$, $R_{\pm }(cc;ss)=(\cosh \theta _{h}\cosh \theta
_{H}+\lambda \sinh \theta _{h}\sinh \theta _{H})_{\pm }.$ The coefficients $%
S_{hj}^{(1,2)}$ are equal to

\begin{equation}
S_{h\pm }^{(1)}=\frac{\gamma _{B}f_{S}}{2\kappa _{h\pm }}\sin \varphi ;\text{
\ }S_{h\pm }^{(2)}=\frac{\gamma _{B}f_{S}}{2\kappa _{h\pm }}\cos \varphi
\label{Sh}
\end{equation}

Eqs. (\ref{f3H}-\ref{Sh}) determine the SRC in three-layer Josephson
junction with different exchange energies in the middle ($H$-region) and
terminal ($h$-region) F layers. We use Eqs.(\ref{f3H}-\ref{f0H}) for the
calculation of the Josephson current due to the SRC.

Let us turn to the calculation of the LRTC. We write the equation for the
matrix $\hat{F}_{1}$ in the $h$-region projecting of Eq.(\ref{Usadel}) on
the $\hat{\sigma}_{1}$ matrix in the spin space

\begin{equation}
\partial ^{2}\hat{F}_{1}(x)/\partial x^{2}-\kappa _{\omega }^{2}\hat{F}%
_{1}(x)=-\kappa _{h}^{2}\sin \alpha \cdot \hat{f}_{3}(x),  \label{EqF1}
\end{equation}%
where the function $\hat{f}_{3}(x)$ in R.H.S is given by Eq.(\ref{f3h}).

The solution of Eq. (\ref{EqF1}) can readily be obtained (see Appendix). In
the $H$-region the function $\hat{F}_{1}(x)$ obeys the same equation but
without the R.H.S. ($\alpha =0$). The solution in this region has the form
of Eq. (\ref{f1}),

\begin{equation}
\hat{F}_{1}=\hat{A}_{1H}\cosh (\kappa _{\omega }x)+\hat{B}_{1H}\sinh (\kappa
_{\omega }x).  \label{F1H}
\end{equation}

In order to calculate the Josephson current $I_{J}$ we need to know the
coefficients $\hat{A}_{1H}$ and $\hat{B}_{1H}.$ Considering the symmetric
case, $\alpha _{r}=\alpha _{l}=\alpha $, we obtain (see Appendix)%
\begin{equation}
\hat{A}_{1H}=\hat{\tau}_{2}\mathcal{A}_{1H}\cos \varphi ,\quad \hat{B}_{1H}=%
\hat{\tau}_{1}\mathcal{B}_{1H}\sin \varphi  \label{a2}
\end{equation}%
where

\begin{eqnarray}
&&\mathcal{A}_{1H}\sinh \theta _{\omega }=i\sin \alpha
\sum_{j=1,2}(-1)^{j+1}S_{hj}^{(2)}  \nonumber \\
&&\times \{\frac{\kappa _{hj}}{\kappa _{\omega }}[\cosh \theta _{h\omega }%
\frac{\sinh \theta _{Hj}}{R_{j}(cs;sc)}-1]+\sinh \theta _{h\omega }\frac{%
\lambda _{j}\cosh \theta _{Hj}}{R_{j}(cs;sc)}\};\text{ }  \nonumber \\
&&  \label{aH} \\
&&\mathcal{B}_{1H}\cosh \theta _{\omega }=i\sin \alpha
\sum_{j=1,2}(-1)^{j+1}S_{hj}^{(1)}  \nonumber \\
&&\times \{\frac{\kappa _{hj}}{\kappa _{\omega }}[\cosh \theta _{h\omega }%
\frac{\cosh \theta _{Hj}}{R_{j}(cc;ss)}-1]+\sinh \theta _{h\omega }\frac{%
\lambda _{j}\sinh \theta _{Hj}}{R_{j}(cs;sc)}\};\text{ }  \nonumber \\
&&  \label{bH}
\end{eqnarray}%
where $\theta _{\omega }=\kappa _{\omega }L,$ $\theta _{h\omega }=\kappa
_{\omega }L_{h},$ $\theta _{Hj}=\kappa _{Hj}L_{H}$ ($\kappa _{1,2}=\kappa
_{\pm }$). The functions $R_{j}(cs;sc)$ and $R_{j}(cc;ss)$ are defined in
Eqs. (\ref{CH}-\ref{SH}).

In the limit of thin $h$-layers ($|\theta _{h\pm }|\ll 1$), we see that the
products $\mathcal{A}_{1H}\sinh \theta _{\omega }$ and $\mathcal{B}%
_{1H}\cosh \theta _{\omega }$ agree with the coefficients $A_{1}$ and $B_{1}$
in Eq. (\ref{A1B1}). This means that the amplitude $F_{1}$ of the LRTC goes
to zero at $L_{h}\rightarrow 0$. On the other hand, it is seen from Eqs.(\ref%
{aH}-\ref{bH}) that at $|\theta _{h\pm }|\gg 1$, the functions $R_{j}$ are
exponentially large and therefore the amplitude of the LRTC decreases with
increasing the thickness $L_{h}$.

\section{Josephson current\label{current}}

Using the Green's functions, $\hat{f}_{0,1,3}$ obtained in Sections \ref%
{thin} and \ref{arbitrary} one can now calculate the dc Josephson current.
Substituting the expansion (\ref{fMatrix}) into Eq.(\ref{JosCurrent}), we
obtain for the Josephson current density

\begin{equation}
j_{J}=i\sigma \pi T\sum_{\omega \geq 0}Tr\{\hat{\tau}_{3}[\hat{f}%
_{0}\partial \hat{f}_{0}/\partial x+\hat{f}_{3}\partial \hat{f}_{3}/\partial
x+\hat{f}_{1}\partial \hat{f}_{1}/\partial x]\},  \label{JosCurrent1}
\end{equation}

The first two terms in Eq. (\ref{JosCurrent1}) are the contribution from the
SRC ($\hat{f}_{0,3}$) and the third term is due to the LRTC ($\hat{f}_{1}=%
\hat{\tau}_{3}\hat{F}_{1}$). The part of the Josephson current density which
is caused by the SRC, $j_{cSR}$, can be written also in the form

\begin{equation}
j_{JSR}=i\frac{\sigma }{2}\pi T\sum_{\omega \geq 0}Tr\{\hat{\tau}_{3}[\hat{f}%
_{+}\partial \hat{f}_{+}/\partial x+\hat{f}_{-}\partial \hat{f}_{-}/\partial
x]\},  \label{jJSR}
\end{equation}%
where $\hat{f}_{\pm }=\hat{f}_{0}\pm \hat{f}_{3}$ are the diagonal matrix
elements in spin space. The part of the Josephson current density which is
caused by the LRCT, $j_{cLR}$, can be presented as

\begin{equation}
j_{JLR}=-i\sigma \pi T\sum_{\omega \geq 0}Tr\{\hat{\tau}_{3}[\hat{F}%
_{1}\partial \hat{F}_{1}/\partial x]\},  \label{jJLR}
\end{equation}

We use these formulas for the calculation of the critical Josephson current
in the limiting cases a) and b).

\subsection{Thin F' Layers}

Consider the limit of thin F' layers. Substituting Eqs.(\ref{fpm>0}) into
Eq.(\ref{jJSR}), we obtain

\begin{equation}
j_{JSR}=i\frac{\sigma }{2}\pi T\sum_{\omega \geq 0}Tr\{\hat{\tau}_{3}[\kappa
_{+}\hat{A}_{+}\hat{B}_{+}+\kappa _{-}\hat{A}_{-}\hat{B}_{-}]\},
\label{jJSR1}
\end{equation}

With the help of Eq.(\ref{Bpm}) this expression can written in the form

\begin{eqnarray}
j_{JSR} &=&j_{cSR}\sin (2\varphi ),\text{ }  \label{jcSR} \\
j_{cSR} &=&\sigma \gamma _{B}^{2}\pi T\sum_{\omega \geq 0}f_{S}^{2}[\frac{1}{%
\mathcal{D}_{+}}+\frac{1}{\mathcal{D}_{-}}],
\end{eqnarray}

\bigskip Using Eqs.(\ref{Den}), we find the critical current density due to
the SRC for the P and AP magnetic configuration

\begin{eqnarray}
j_{cSR,P} &=&\sigma \gamma _{B}^{2}2\pi T\sum_{\omega \geq 0}f_{S}^{2}Re(%
\frac{1}{\kappa _{+}\sinh (2\theta _{H+})}),  \label{jcPar} \\
j_{cSR,AP} &=&\sigma \gamma _{B}^{2}2\pi T\sum_{\omega \geq 0}f_{S}^{2}\frac{%
1}{2Re(\kappa _{+}\sinh \theta _{H+}\cosh \theta _{H-})},  \nonumber \\
&&  \label{jcAnti}
\end{eqnarray}

\begin{figure}[tbp]
\begin{center}
\includegraphics[width=6cm, height=5cm]{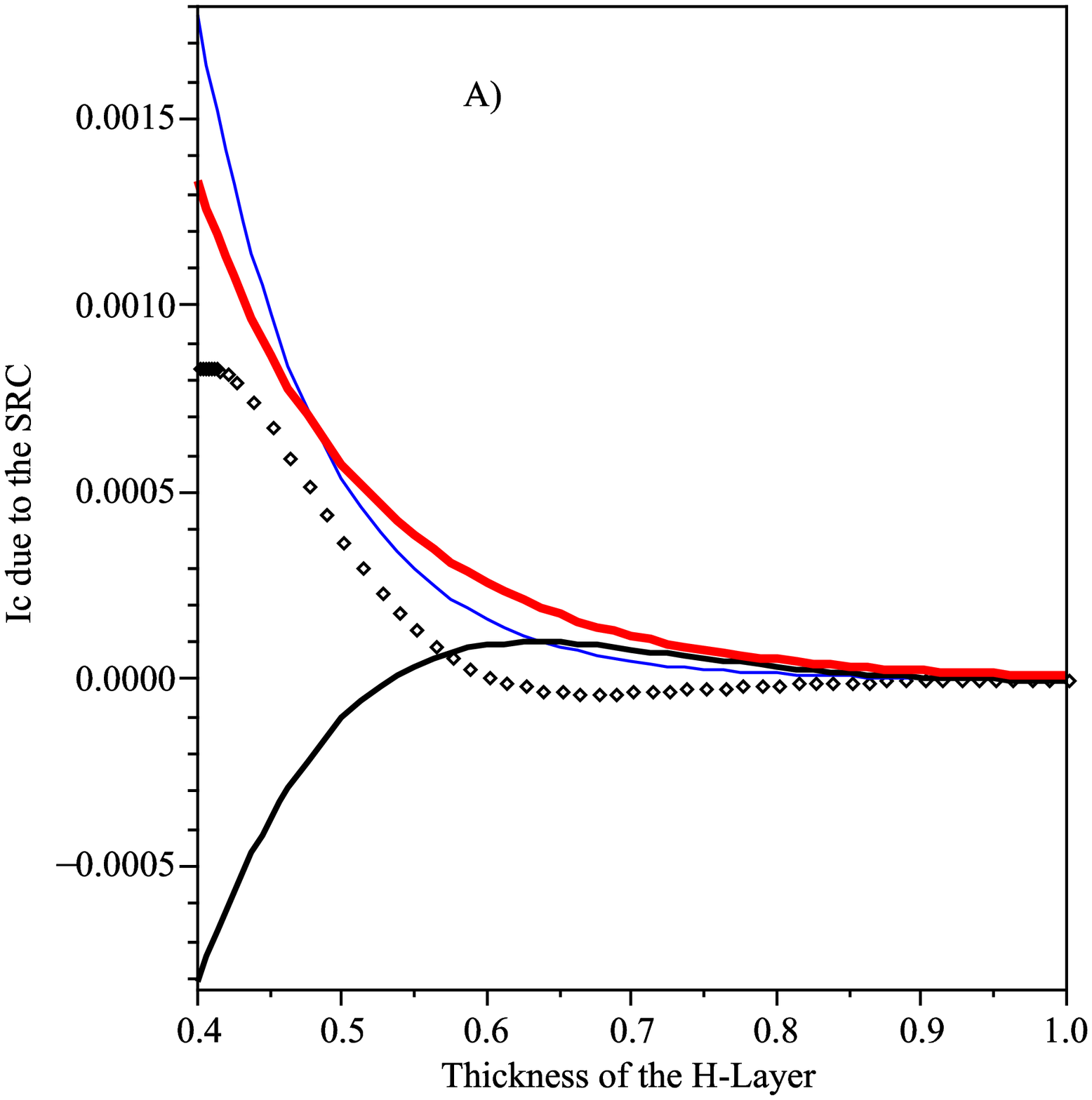} %
\includegraphics[width=6cm, height=5cm]{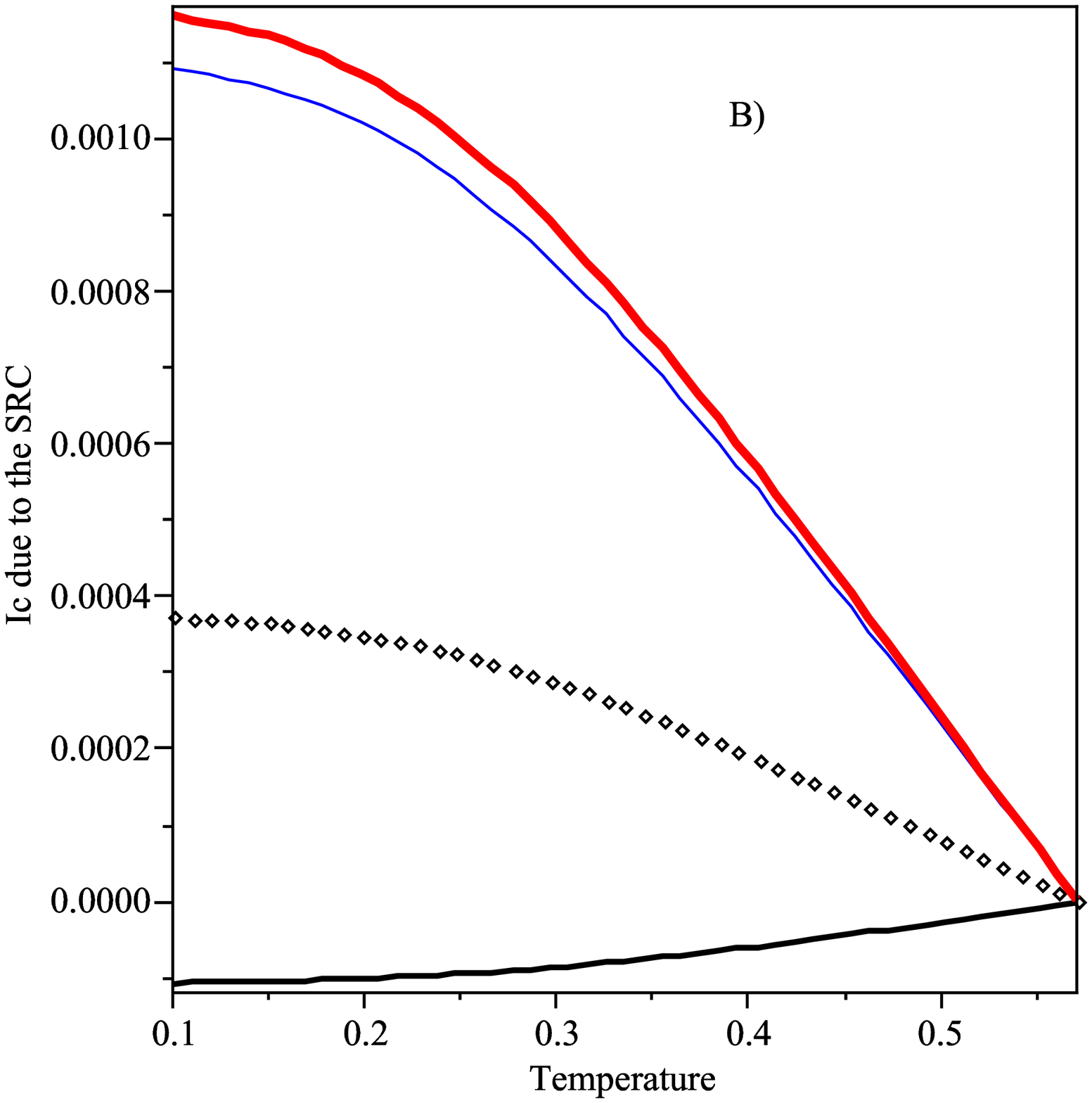}
\end{center}
\caption{(color online) Normalized critical current due to the SR
components, $I_{cSR}$, for the parallel (P) and antiparallel (AP)
orientations of magnetization in domains in the middle $H$-layer vs the
thickness of the $H$-layer (A) and temperature (B). The thickness of the $h$%
-layers is assumed to be small ($L_{h}<<\protect\xi_{h}$). The normalized
temperature $\tilde{T}$, exchange energy $\tilde{H}$ and thickness $\tilde{L}%
_{H}$ are measured in units $\Delta_{0}$ and $\protect\xi_{\Delta}=\protect%
\sqrt{D/\Delta_{0}}$, where $\Delta_{0}=\Delta(T)$ at $T=0$. The lower and
upper curves correspond to the P and AP configuration, respectively. The
critical current $j_{c}$ is measured in units of its value in a junction
with $H=0$ (no exchange field) and normalized length $\tilde{L}_{H}=0.5$.
The point and upper thin curves correspond to $\tilde{H}=70$; the lowest and
upper thick curves correspond to $\tilde{H}=30$. The normalized temperature
(in Fig.A) and thickness $\tilde{L}_{H}$ (in Fig.B) are equal to 0.1 and
0.5; the parameter $\protect\tau _{m}\Delta $ equals 1. The values of the
critical current corresponding to $\tilde{H}=30$ are reduced by 10 times. }
\end{figure}

\begin{figure}[tbp]
\begin{center}
\includegraphics[width=6cm, height=5cm]{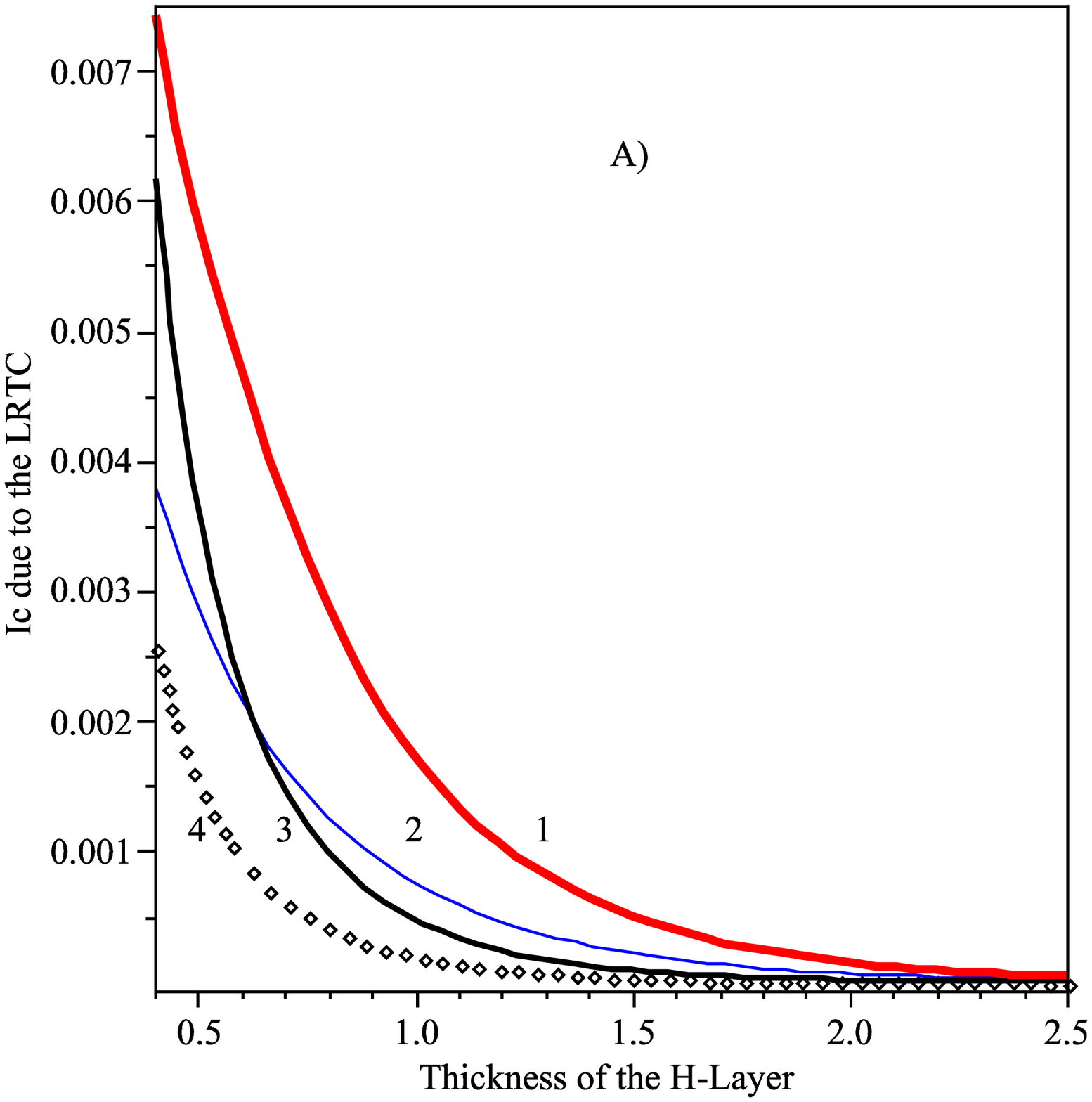} %
\includegraphics[width=6cm, height=5cm]{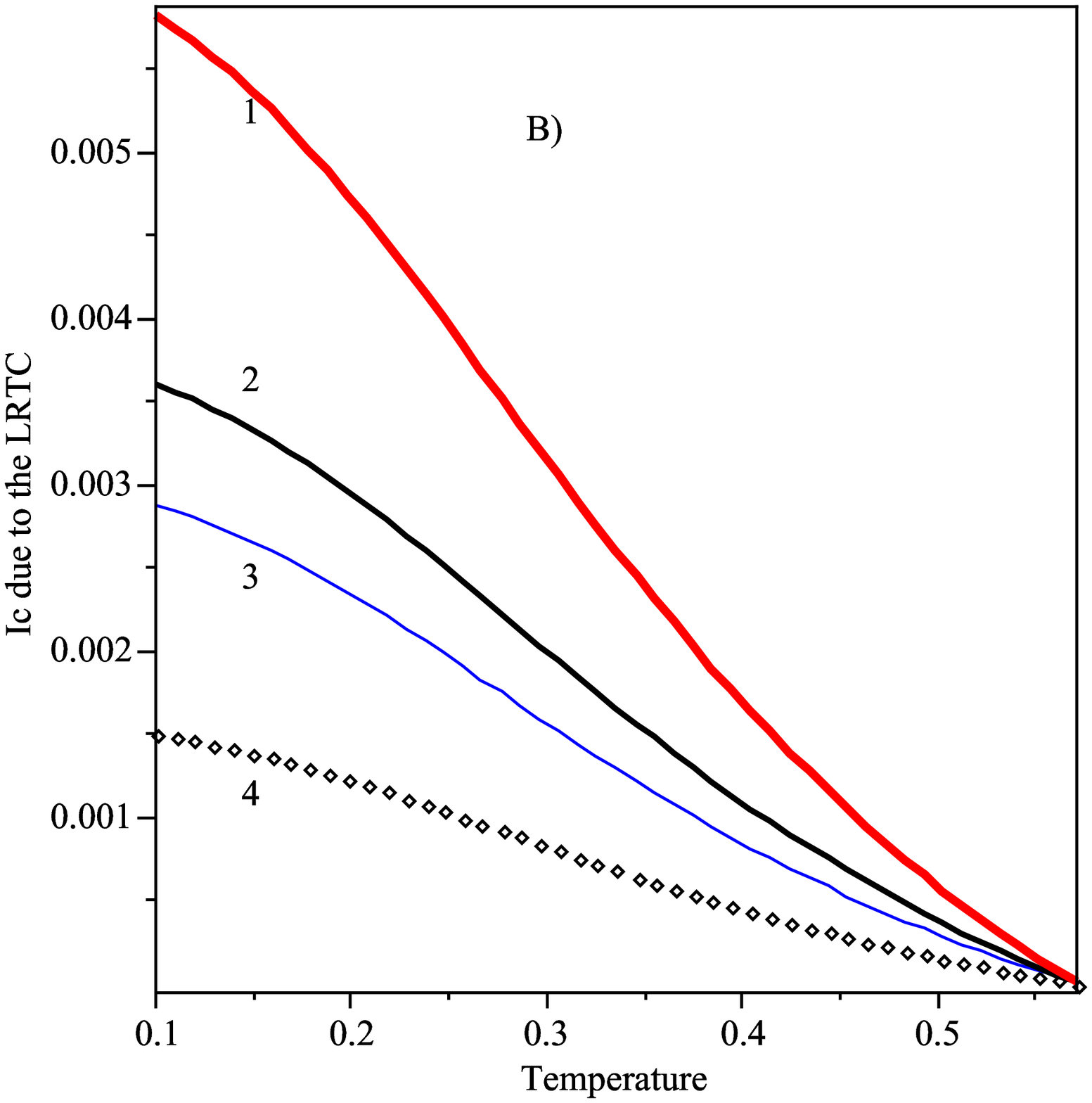}
\end{center}
\caption{(color online) The same dependence as in Fig.2 for the normalized
critical current due the LRTC components, $I_{cLR}$. As in Fig.2, the
thickness of the $h$-layers is assumed to be small ($L_{h}<<\protect\xi _{h}$%
). The normalization units are the same as in Fig.2. The curves are numbered
so that: 1 - AP, ($\tilde{H}=30$); 2 - AP (70); 3 - P (30); 4 - P (70).}
\end{figure}
These formulas show that the critical current density $j_{cSR}$ decays
exponentially with increasing the length $L_{H}$ over a short scale of the
order $\xi _{H}$. The decrease of the current density $j_{cSR,P}$ with
increasing the thickness $L_{H}$ or exchange energy $H$ is accompanied by
oscillations \cite{GolubovRMP,BuzdinRMP,BVErmp}. Oscillations in the
dependence $j_{cSR,AP}(L_{H},H)$ are absent in the case of antiparallel
orientations. The latter behavior was predicted by Blanter and Hekking \cite%
{Blanter} who considered antiparallel orientation of magnetization in
ferromagnetic domains in a SFS Josephson junction and presented formula for $%
j_{cSR,AP}$ in the limit $\theta _{H+}\gg 1.$ Eq.(\ref{jcAnti}) generalizes
Eq.(26) of Ref.\cite{Blanter} for the case of arbitrary $\theta _{H\pm }$,
i.e. arbitrary $L_{H}.$ A rapid decay and oscillations of the critical
Josephson current in SFS or SIFS junctions were observed in many
experimental works \cite%
{Ryaz01,Ryaz06,Kontos,Blum,Bauer,Sellier,Weides06,Weides09,Blamire06,Shelukhin,Westerholt,Birge09}%
. This oscillatory behaviour was predicted a few decades ago \cite%
{Bulaev,BuzdinBul}.

In Fig.2a and 2b we show the dependence of the normalized critical currents
density $I_{cSR}\equiv $ $j_{cSR}(h,H,L_{H})/j_{c}(0,0,L_{H})$ originating
from the short-range component (the components $f_{0,3}$) on the thickness
of the $H$-layer $L_{H}$ and temperature for the P and AP magnetization
orientation. The plots are obtained on the basis of Eqs. (\ref{jcPar}-\ref%
{jcAnti}) applicable in the case of thin $h$-layers. We see that in the case
of the P orientation the critical current density $j_{c,SR}$ caused by the
SRC changes sign with increasing $L_{H},$ while in the case of the AP
orientation the current $j_{c,SR}$ is always positive. However, at a fixed $%
L_{H}$ the current density $j_{cSR}$ decays monotonously with increasing
temperature. For a smaller exchange energy the dependence $j_{cSR}(T)$ has
another form and may change the sign (see the next subsection).

Note that the critical current for the AP orientation $I_{cSR,AP}$ is always
larger that for the P orientation. The critical current $I_{cSR,AP}$ in an
SFIFS Josephson junction with the antiparallel magnetization orientation in
the F layers may be even exceed the critical current in SIS junction without
the F layers provided that the coupling between S and F layers is strong
enough ( here I stands for an insulator) \cite{BVE01a,BVErmp}.

\begin{figure}[tbp]
\begin{center}
\includegraphics[width=6cm, height=5cm]{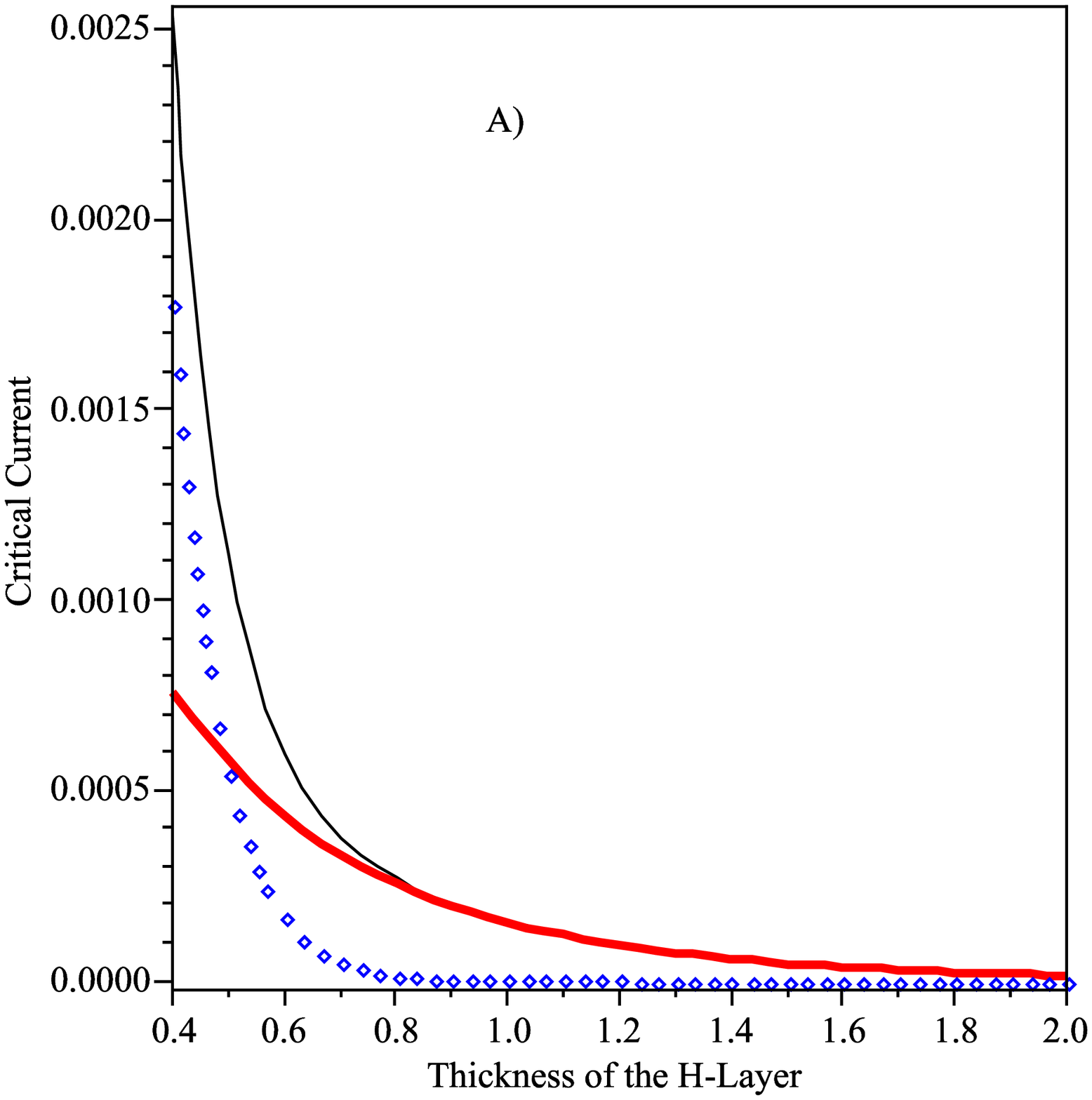} %
\includegraphics[width=6cm, height=5cm]{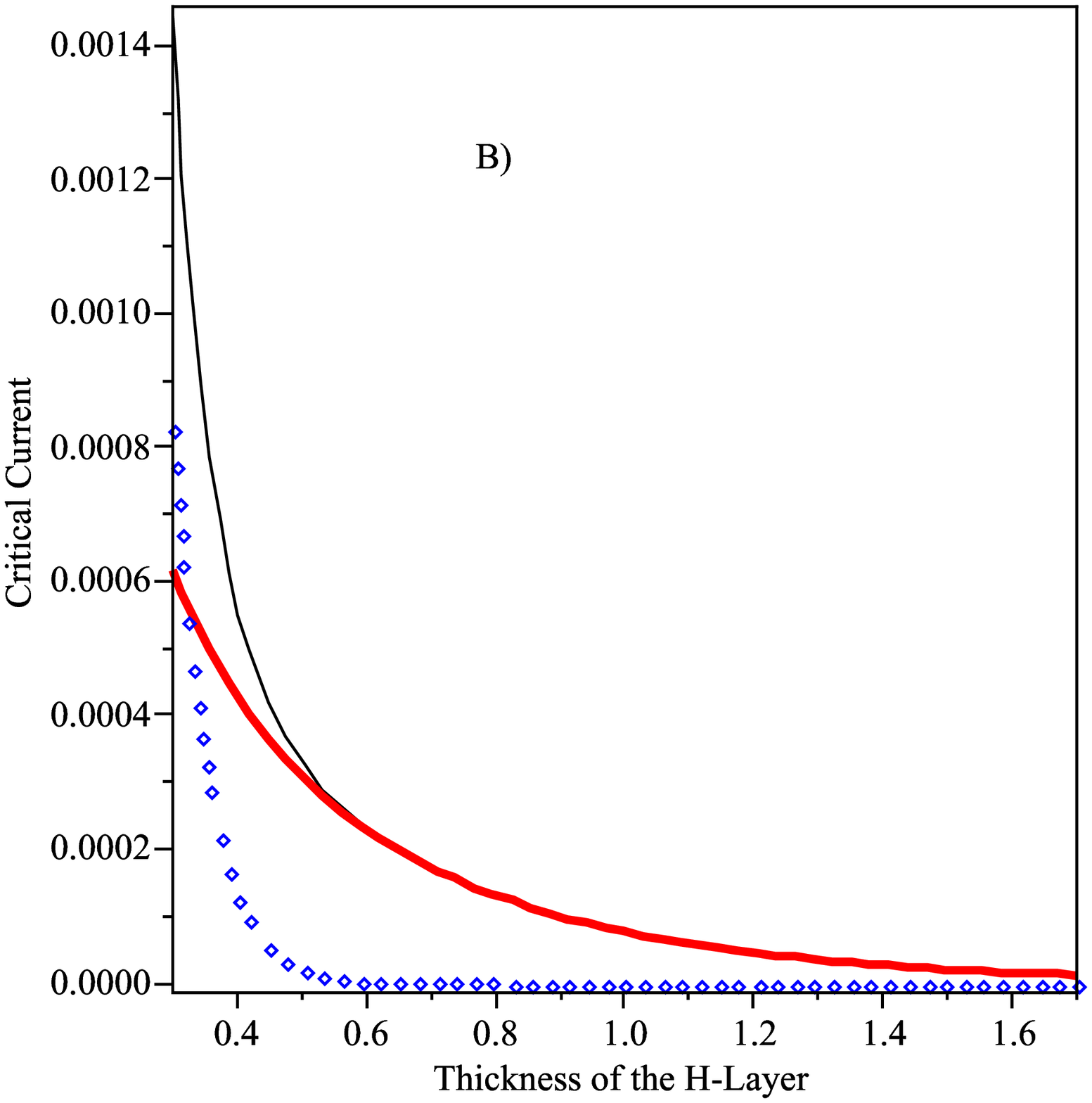}
\end{center}
\caption{(color online) The dependence of the normalized critical current
due to the SR component (point line), LRTC (solid thick line) and the total
critical current (solid thin line)on the thickness of the $H$-layer for $%
\tilde{H}=70$ (A) and $\tilde{H}=170$ (B). The normalized temperature is
equal to 0.1. Other parameters are the same as in Fig.2.}
\end{figure}

The Josephson current $j_{JLR}$ due to the LRTC is found using Eqs. (\ref%
{jJLR}, \ref{f1},\ref{A1B1}). We obtain

\[
j_{JLR}=i\sigma \pi T\gamma _{1}^{2}\sin \alpha _{r}\sin \alpha
_{l}\sum_{\omega \geq 0}\frac{Tr\{\hat{\tau}_{3}\hat{f}_{3}(L_{H})\hat{f}%
_{3}(-L_{H})\}}{\kappa _{\omega }\sinh (2\theta _{H\omega })}
\]

\begin{equation}  \label{jJLRthin}
\end{equation}

The matrices $\hat{f}_{3}(\pm L_{H})$ are expressed in terms of \ the
matrices $\hat{f}_{\pm }(\pm L_{H}):$ $\hat{f}_{3}(\pm L_{H})=[\hat{f}%
_{+}(\pm L_{H})-\hat{f}_{+}(\pm L_{H})]/2.$ Using Eqs.(\ref{fpm>0}-\ref%
{fpm<0}), we find that $j_{JLR}=j_{cLR}\sin (2\varphi )$. The critical
current $j_{cLR}$ in the case of the P and AP orientations is given by

\begin{eqnarray}
&&j_{cLR,P}=-4\sigma \gamma _{B}^{2}\gamma _{1}^{2}\sin \alpha _{r}\sin
\alpha _{l}(2\pi T)  \nonumber \\
&&\times \sum_{\omega \geq 0}\frac{f_{S}^{2}}{\kappa _{\omega }\sinh
(2\theta _{H\omega })}ReP_{1}\cdot ReP_{2},  \label{jTRpar} \\
&&j_{cLR,AP}=-\sigma \gamma _{B}^{2}\gamma _{1}^{2}\sin \alpha _{r}\sin
\alpha _{l}(2\pi T)  \nonumber \\
&&\times \sum_{\omega \geq 0}\frac{f_{S}^{2}}{\mathcal{D}_{AP}^{2}\kappa
_{\omega }\sinh (2\theta _{H\omega })}[P_{3}^{2}-1],  \label{jTRanti}
\end{eqnarray}%
where $P_{1}=\cosh ^{2}\theta _{+}/\mathcal{D}_{+},$ $P_{2}=\sinh ^{2}\theta
_{+}/\mathcal{D}_{+},$ $P_{3}=|\cosh \theta _{+}|^{2}(1+|\tanh \theta
_{+}|^{2}Re(\kappa _{+}/\kappa _{-})),$ $\mathcal{D}_{+}$ and $\mathcal{D}%
_{AP}\equiv \mathcal{D}_{AP+}=\mathcal{D}_{AP-}$ are defined in Eq. (\ref%
{Den}). Eq. (\ref{jTRpar}) resembles Eqs. (12) and (10) of Refs.\cite%
{BVE03,BuzdinHouzet}, respectively. Comparing Eq. (\ref{jTRpar}) with
Eq.(10) of Ref.\cite{BuzdinHouzet} one has to have in mind that other
boundary conditions and temperatures close to $T_{c}$ are considered in that
work. Note that the parameter $(\gamma _{1}/\kappa _{\omega })=(\kappa
_{h}L_{h})(\kappa _{h}/\kappa _{\omega })$ may be arbitrary; we assumed that
$\kappa _{h}L_{h}\ll 1$, but the parameter $(\kappa _{h}/\kappa _{\omega })$
is large.

In Figs. 3a and 3b we plot the same dependencies as in Figs. 2a and 2b for
the critical current $j_{cLR}$ originating from the LRTC. This current
decays monotonously with increasing both the temperature and $L_{H}$. Note
that the critical current $j_{cLR}$ has the opposite sign to $j_{cSR,AP}$ if
$\sin \alpha _{r}\sin \alpha _{l}>0$ and in the case of the AP configuration
is always larger than in the case of the P configuration.

In Figs. 4a and 4b we plot the most important dependencies of the $%
j_{cSR}(T) $ and $j_{cLR}(T)$ as well as the total critical current, $%
j_{cTot}=$ $j_{cSR}(T)+$ $|j_{cLR}(T)|,$ on the length $L_{H}$ for different
magnitudes of the exchange energy $H$. The parameter $p(\alpha _{r},\alpha
_{l})\equiv 4\gamma _{1}^{2}\xi _{\Delta }^{2}\sin \alpha _{r}\sin \alpha
_{l}$ is taken to be equal to $0.2.$ The value of $H=170\Delta _{0}$
approximately corresponds to the exchange energy in $Co$ ($E_{ex}=309meV,$
see \cite{Blamire06}) used in the experiment \cite{Birge}. One can see that
for $H=70\Delta _{0}$ and $H=170\Delta _{0}$ the critical current is caused
by the LRTC at $L_{H}\gtrsim 0.5\xi _{\Delta }$ and at $L_{H}\gtrsim 0.4\xi
_{\Delta }$, respectively, where $\xi _{\Delta }=\sqrt{D/\Delta _{0}}$. The
latter curve is close to the one observed in Ref.\cite{Birge} provided we
accept $\xi _{\Delta }=10nm$ corresponding to the value of the diffusion
coefficient $D$ about $10cm^{2}/s$.

One can say that the obtained results are in a qualitative agreement with
experimental data \cite{Birge}. It is difficult to carry our the
quantitatvie comparison because our model is simplified. We admit the
standard model in which one neglects the difference in the diffusion
coefficients for the major and minor electrons in the ferromagnetic layers
although this difference in $Co$ is large. We also assume the diffusive
limit for all layers. At the same time, the mean free path in the strong
ferromagnet ($Co$) seems to be larger than $\xi _{H}$. In this case the
formula for the Josephson current even for one-layer SFS junction becomes
rather complicated\cite{BVECrCur}.

In addition, the conductivities in all the layers are assumed to be equal
whereas in experiment the conductivities in layers ($Cu,Co$\ and $PdNi$) are
different. The interface resistance $R_{B}$, strictly speaking, is not known
and can be only estimated. However, one can see from Eqs.(\ref{jcPar}-\ref%
{jcAnti}) that both the ''effective conductivity'' $\sigma $ (averaged over
all layers) and the averaged interface resistance $R_{B}$\ enter the
expression for the critical current density $j_{c}$\ as a prefactor before
the sum. It disappears when we plot the critical current normalized to its
value at $h=H=0.$ The most interesting dependence of the critical current $%
j_{c}$\ on temperature and thicknesses $L_{h,H}$\ is determined by
exponential functions.

We assumed that the proximity effect is weak, that is, the amplitude of the
condensate functions in ferromagnets is small.\ As follows from Eq.(\ref{Bpm}%
), this means that the parameter $|\gamma _{B}/\kappa _{+}|\approx |\rho
L_{h,H}/R_{B}\theta _{+}|\approx |\rho L_{h,H}/R_{B}|\ll 1.$\ According to
Ref.\cite{Birge09}, where a structure similar to that in Ref.\cite{Birge}
(but without strong ferromagnets) was studied, the interface resistance per
unit area, $R_{B}=SR_{SF},$\ was equal $R_{B}=2.3\cdot 10^{-8}\Omega cm^{2}$%
\ whereas $R_{F}S=\rho _{F}L_{F}\approx 7\cdot 10^{-12}\Omega cm^{2},$ that
is, the ratio $R_{F}S/R_{B}\approx 10^{-4}$ is indeed very small, here $S$
is the cross section area of the junction$.$\ Taking into account that the
resistance of the $h$\ and $H$\ ($Co$) - layers are comparable, we conclude
that the proximity effect in experiment \cite{Birge} is weak.

As to the value of the critical current, we do not attempt to carry out a
quantitative comparison with experimental value because it depends on the
ratio $(\rho L/R_{B})^{2}=(R_{F}S/R_{B})^{2}$\ which is known only on the
order of magnitude. In addition to that, the Josephson junction used in
experiment contains many interfaces each of which reduces the proximity
effect.

\subsection{Arbitrary Thicknesses of the F Layers}

In order to calculate the current density $j_{JSR}$, we substitute Eqs.(\ref%
{f3H}-\ref{Sh}) into Eq.(\ref{jJSR}) using the relations $\hat{f}_{0,3}(x)=[%
\hat{f}_{+}(x)\pm \hat{f}_{+}(x)]/2.$ After simple calculations, we obtain
for the critical current density $j_{cSR}$ in the case of the P configuration

\begin{equation}
j_{cSR}=\sigma (2\pi T)\gamma _{B}^{2}\sum_{\omega \geq 0}Re[\frac{f_{S}^{2}%
}{\kappa _{H+}R(cs/sc)_{+}R(cc/ss)_{+}}]  \label{jcSRAR}
\end{equation}

The critical current density due to the LRTC, $j_{cLR},$ for the case of the
arbitrary thicknesses $L_{h,H}$ is found from Eqs.(\ref{F1H}-\ref{bH},\ref%
{jJLR}). For the symmetric case ($\alpha _{r}=\alpha _{l}\equiv \alpha \ll 1$%
) we get

\begin{equation}
j_{cLR}=-\sigma (2\pi T)\gamma _{B}^{2}\alpha ^{2}\sum_{\omega \geq 0}\frac{%
f_{S}^{2}}{\kappa _{\omega }\sinh (2\theta _{\omega })}Im(M)Im(N)
\label{jcLRAR}
\end{equation}%
where

\begin{eqnarray}
&&M=\frac{\sinh \theta _{H+}\cosh \theta _{h\omega }}{R(cs/sc)_{+}}+\frac{%
\kappa _{\omega }\cosh \theta _{H+}\sinh \theta _{h\omega }}{\kappa
_{H+}R(cs/sc)_{+}},  \nonumber \\
&&  \label{M} \\
&&N=\frac{\cosh \theta _{H+}\cosh \theta _{h\omega }}{R(cc/ss)_{+}}+\frac{%
\kappa _{\omega }\sinh \theta _{H+}\sinh \theta _{h\omega }}{\kappa
_{H+}R(cc/ss)_{+}},  \nonumber \\
&&  \label{N}
\end{eqnarray}%
and $\theta _{\omega }=\theta _{\omega h}+\theta _{\omega H}$. In the
antisymmetric case ($\alpha _{r}=-\alpha _{l}\equiv \alpha \ll 1$) the sign
of the critical current density should be changed. Thus, in this case the
critical current has the same sign as in SNS junction.

Using Eqs.(\ref{jcSR}-\ref{N}), we calculate numerically the critical
currents $j_{cSR}$ and $j_{cLR}$ for the case of symmetric case ($\alpha
_{r}=\alpha _{r}\equiv \alpha \ll 1$) and the P configuration. In Fig. 5a
and 5b the dependence of $j_{cSR}$ and $j_{cLR}$ on temperature $T$ is shown
for the case when the exchange energies are not very high ($H=2h=10\Delta
_{0}$) and the lengths $L_{H,h}$ are small ($L_{h}\approx L_{H}=0.1\xi
_{\Delta }$). It is seen that the current $j_{cSR}$ depends on $T$ in a
non-monotonic way and can change sign. This type of a non-monotonic
temperature dependence was obtained in many works, both experimental \cite%
{Ryaz01} and theoretical (see reviews \cite{GolubovRMP,BuzdinRMP} and
references therein as well as a recent paper \cite{SudboJJ}). The critical
current due to the LRTC $j_{cLR}$ decays with increasing $T$ monotonously.
Note that a non-monotonic behavior of $j_{cLR}(T)$ was found in Ref. \cite%
{EschrigRev} for a half-metallic ferromagnet. In our model, we do not find
values of parameters at which this dependence would not be monotonic.
However, there is no contradiction between these two results because, as was
shown recently \cite{EschrigPRL09}, the non-monotonic temperature dependence
of the critical Josephson current $I_{cLR}$\ takes place only for a
sufficiently large exchange energy (comparable with the Fermi energy $%
\epsilon _{F}$). In our study we assume that both exchange energies, $h,H$,
are small in comparison with $\epsilon _{F}$. In a recent work \cite%
{Westerholt10}, the Josephson dc effect was observed in a SFS junction with
ferromagnetic $Cu_{2}MnAl$-Heusler alloy as a ferromagnetic layer. In the
interval of thicknesses $8nm\lesssim L_{F}\lesssim 10.5nm$\ the critical
current as a function of $L_{F}$\ shows a very slow decay with increasing $%
L_{F},$\ and its temperature dependence was a non-monotonic in this interval
of $L_{F}.$\

In Fig.6A we show the dependence of $j_{cLR}(L_{h})$ for different exchange
energies $h$ setting the angle $\alpha $\ equal to $\sqrt{0.1}$. It is seen
that the critical current caused by the LRTC has a maximum at $L_{h}\approx
\xi _{h},$ that is, the maximum shifts to smaller $\xi _{h}$ with increasing
$h$. The same dependence $j_{cLR}(L_{h})$, but for different $H$, is shown
in Fig.6B. One can see that the position of the maximum weakly depends on $%
H. $ A nonmonotonic behaviour of the critical current as a function of $L_{H}
$ was observed in experiment \cite{Birge}.

\begin{figure}[tbp]
\begin{center}
\includegraphics[width=6cm, height=5cm]{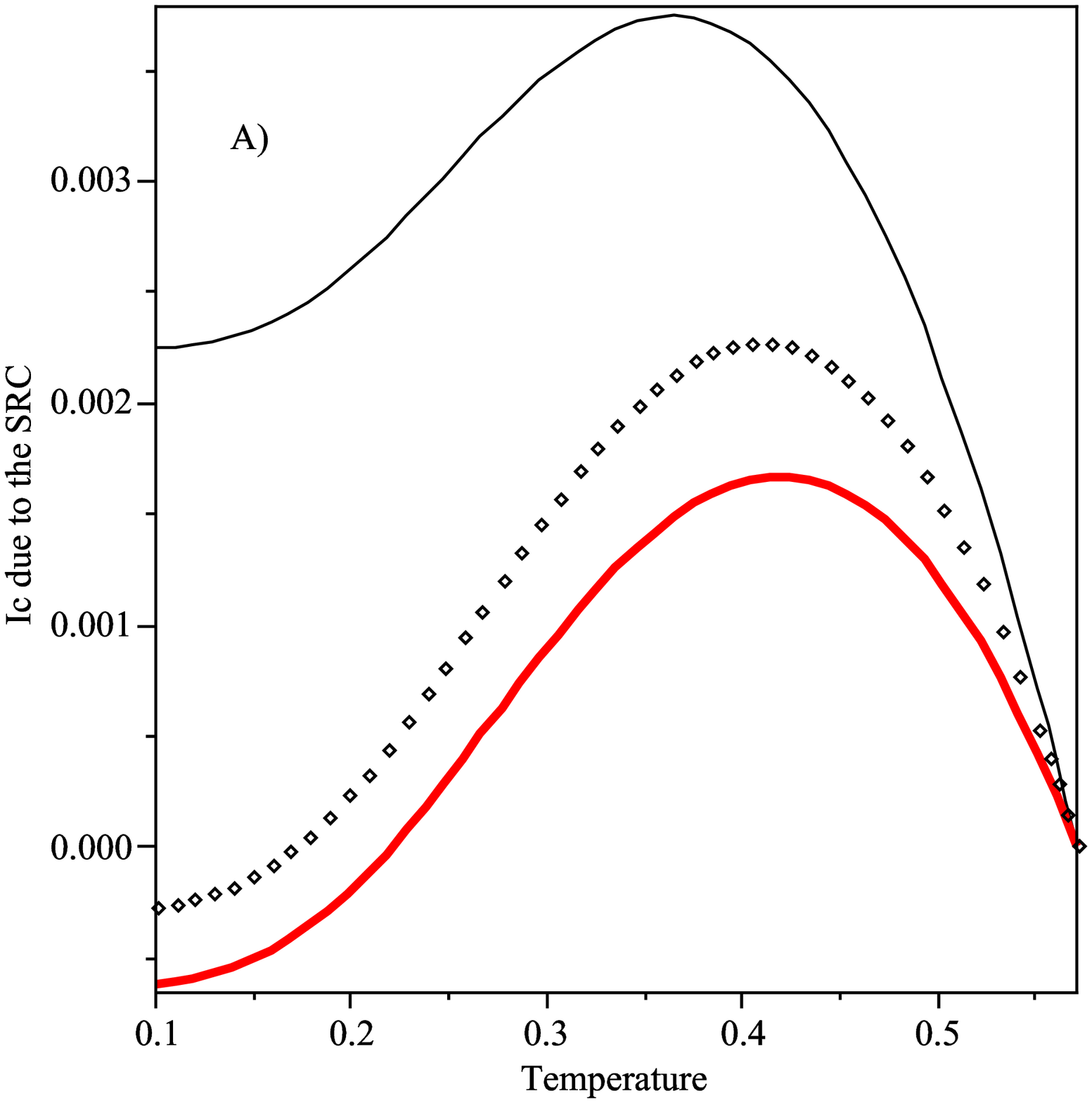} %
\includegraphics[width=6cm, height=5cm]{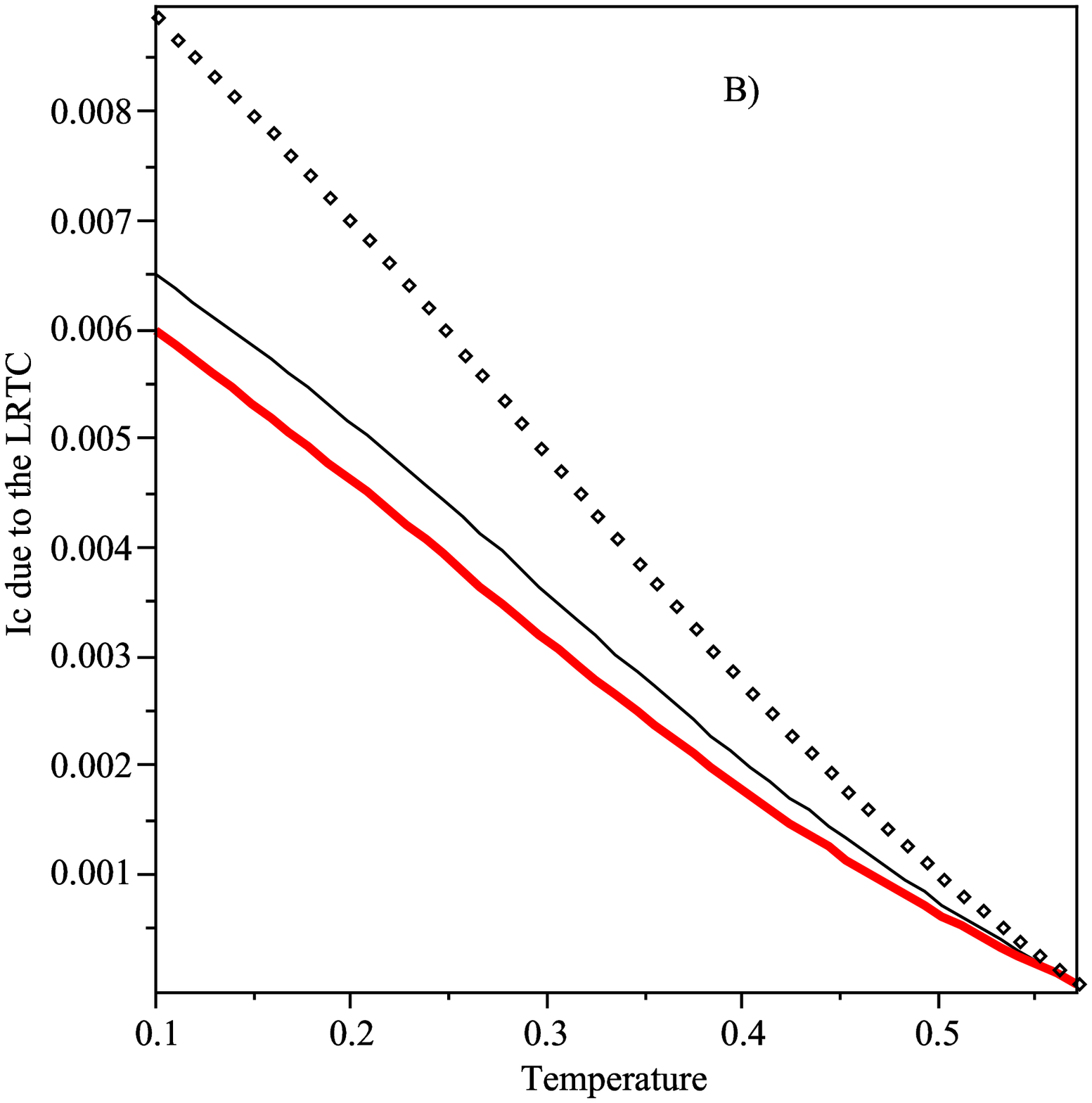}
\end{center}
\caption{(color online) Temperature dependence of the normalized critical
current due to the SR (A) and LRTC (B) components for the P orientation. The
upper and lower curves correspond to $\tilde{h}=5$, $\tilde{H}=10$ and to
normalized lengths $\tilde{L}_{h}=0.1$ and $\tilde{L}_{h}=0.12$. The point
curve corresponds to $\tilde{h}=5, \tilde{H} =10$ and $\tilde{L}_{h}=0.1$.
The "length" $\tilde{L}_{H}=0.1$ and the parameter $\protect\tau%
_{m}\Delta_{0}=0.1$.}
\end{figure}

\begin{figure}[tbp]
\begin{center}
\includegraphics[width=6cm, height=5cm]{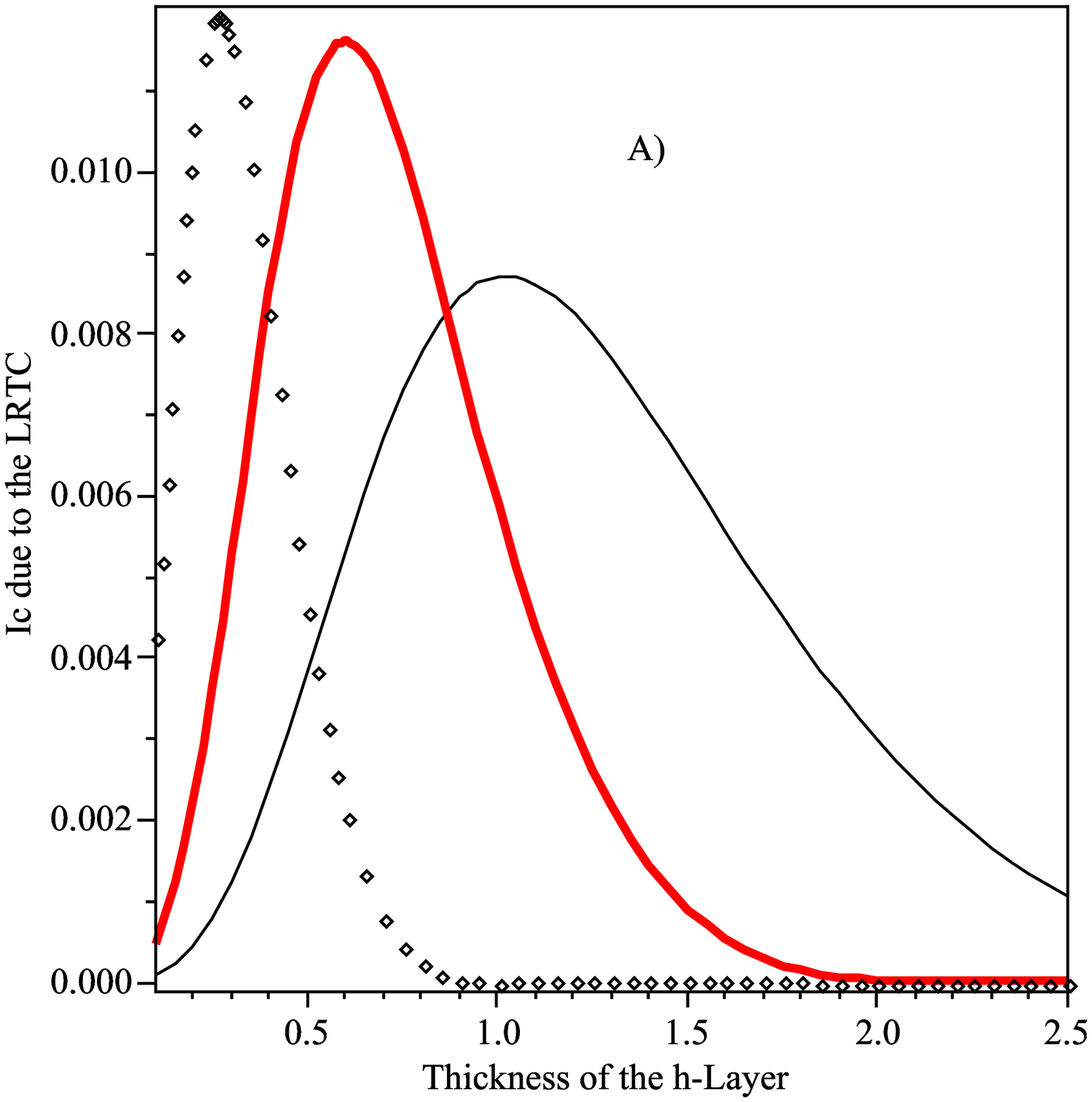} %
\includegraphics[width=6cm, height=5cm]{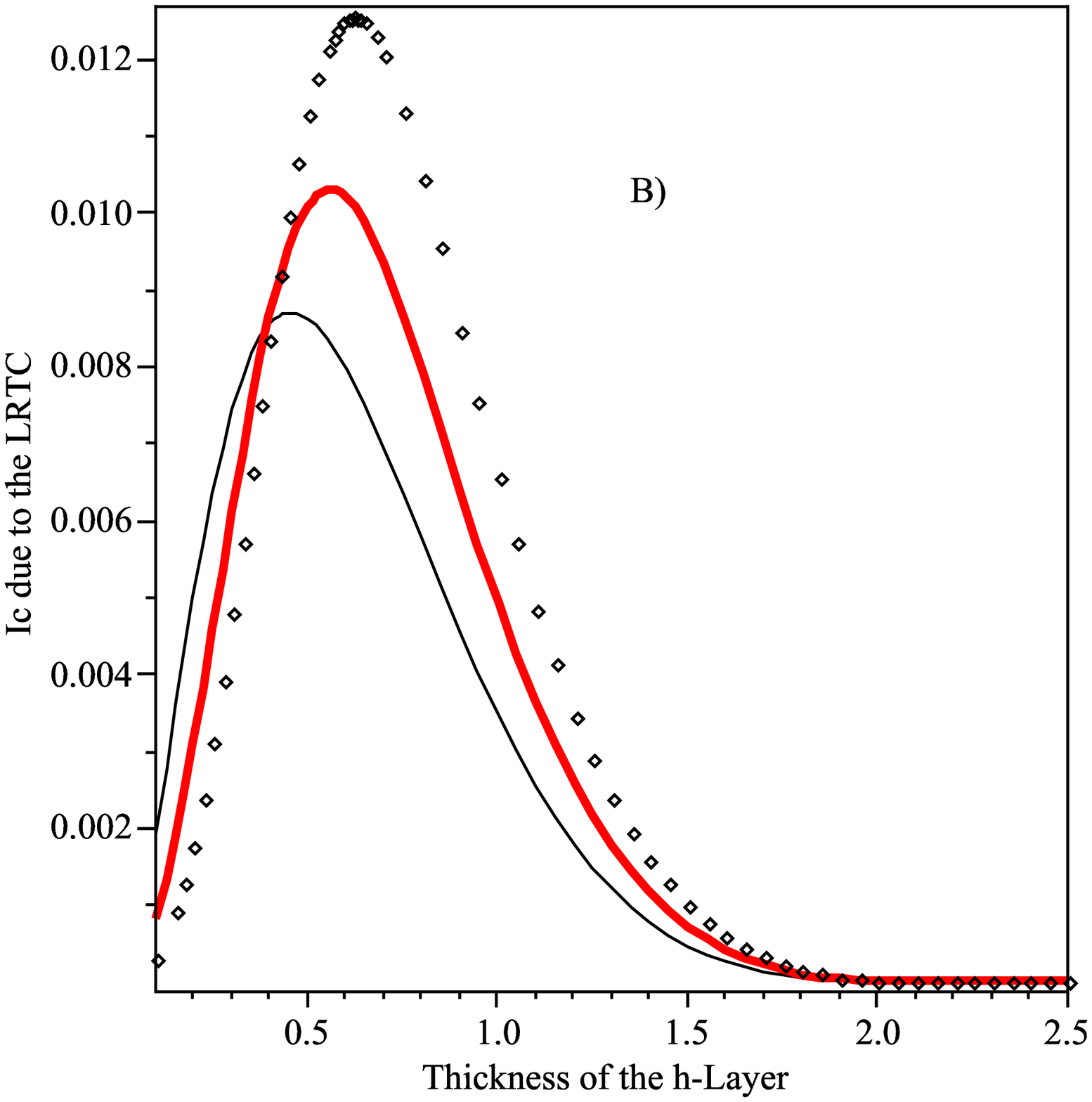}
\end{center}
\caption{(color online) A) Normalized critical current due to the LRTC as a
function of the thickness of the $h$-layer for different exchange energies $%
h $ for the P orientation. The parameters are: $\tilde{h}=2,5,20$ (solid
thin, solid thick and point lines, respectively). The other parameters are $%
\tilde{H}=50,\tilde{T}=0.25, \protect\tau _{m}\Delta =0.1, \tilde{L}_{H}=0.5$%
. B) The same dependence for different $\tilde{H}$: $\tilde{H}=5$ (thin
solid curve), $\tilde{H}=20$ (thick solid curve) and $\tilde{H}=100$ (point
curve). The parameter $\tilde{h}=5$. Other parameters are the same as in
Fig.6A. All the curves correspond to the P orientation.}
\end{figure}

\section{Discussion\label{discussion}}

We have considered the long-range triplet component in an SF'FF'S diffusive
Josephson junction with a non-collinear magnetization $M$ orientation in the
F' and F ferromagnetic layers. Assuming that the proximity effect is weak,
we have solved the linearized Usadel equation and found the pair wave
functions for the short-range (singlet and $S=0$ triplet) and long-range
components (the LRTC with $|S|=1$) in the cases of different exchange fields
in F' and F layers, arbitrary temperatures and parallel (antiparallel) $M$
orientation in two domains in the middle F layer.

Our study was motivated by recent experimental results concerning the
observation of the LRTC in a multilayered (seven layers between
superconductors) Josephson junctions \cite{Birge}. The model used in our
study, although accounting for the properties of the SFS junction used in
the experiment \cite{Birge}, is somewhat simplified. In particular, we solve
the Usadel equation in which the difference in transport properties of the
minor and major electrons in the F layers was ignored. This approximation is
rather crude especially for strong ferromagnets (in $Co$ the diffusion
coefficients for the minor and major electrons may differ by an order of
magnitude \cite{Birge}). Account for different transport properties for
electrons with up and down spins leads to a change not only in the boundary
condition (\ref{BC}) \cite{Belzig+Naz} but also in the collision integral by
potential impurities \cite{BVE02}. The scattering by stationary fluctuations
of magnetic moments in the ferromagnetic layers is also taken into account
in the simplest way ignoring the spin-orbit interaction \cite{Demler}.

Thus, the approximations taken by us do not allow a quantitative comparison
of the critical current $j_{c}$\ measured in experiment and calculated on
the basis of the quasiclassical theory within a simplified model. \ The
formulas for $j_{c}$, Eqs.(\ref{jcPar}-\ref{jcAnti}), contain an effective
conductivity averaged over all layers and effective interface resistance.
Even in a simpler case of SFS junctions with a single ferromagnetic layer it
is not possible yet to obtain a quantitative agreement between theory based
on the quasiclassical theory and experiment \cite{Blamire06,Birge09}.

Fortunately, both these parameters enter the corresponding formulas as
prefactors which disappear when the critical current $%
I_{c}(T,h,H,L_{h},L_{H})$ is normalized to its value at $h=H=0$. Therefore,
the performed study allows one to understand what kind of dependencies of
the critical current $I_{c}$ on different parameters ($T,L_{H},L_{h},H,h$)
can be obtained in multilayered Josephson SFS junction. If the magnetization
in the domains in the F layer is aligned parallel, the critical current $%
I_{cSR}$ caused by the short-range components oscillates with increasing the
thicknesses of the F' and F layers ($L_{h,H}$). This behavior was predicted
in Refs.\cite{Bulaev,BuzdinBul} and observed in many experimental works \cite%
{Ryaz01,Ryaz06,Kontos,Blum,Bauer,Sellier,Blamire06,Weides06,Weides09,Westerholt,Birge09}%
. The formulas obtained in this paper generalize previous theoretical
results for $I_{cSR}$ (see reviews \cite{GolubovRMP,BuzdinRMP,BVErmp} and
references therein) to the case of different exchange fields in the F' and F
films.

If the magnetization in domains in the F layer are antiparallel, the
critical current $I_{cSR}$ decays exponentially in agreement with the
results of Ref.\cite{Blanter}. In both the cases of the parallel and
antiparallel orientations the characteristic length for the decay of $%
I_{cSR} $ is of the order of $\xi _{h,H}$.

The critical current $I_{cLR}$ due to the LRTC decays in both cases with
increasing $L_{H}$ and $T$ in a monotonic way on the length of the order of $%
\xi _{N}.$ For certain values of parameters of the system the total critical
current $I_{c}=$ $I_{cSR}+I_{cLR}$ coincides with $I_{cSR}$ at small
thickness of the F layer $L_{H}$ and with $I_{cLR}$ at larger $L_{H}$ (see
Fig.4). This behavior agrees with the dependence $I_{c}(L_{H})$ observed in
the experiment \cite{Birge}.

As it was predicted in Ref.\cite{BVE03} for a F'SFSF' system and in Ref.\cite%
{BuzdinHouzet} for a SF'FF'S system, the current $I_{cLR}$ has a maximum as
a function of $L_{h}$ at $L_{h}\sim \xi _{h}$ decaying to zero at large and
small $L_{h}$. Thus, the presence of the the F' layers makes the F layer
with a strong ferromagnet \textquotedblleft transparent\textquotedblright\
for the LRTC. Unlike previous theoretical studies, we present a formula for
the amplitude of the current $I_{cLR}$ for arbitrary thicknesses $L_{h,H}$,
temperatures and exchange fields $h,H$. This allows one to choose optimal
parameters for observing the LRTC.

However, it remains unclear how the angle dependence of the critical $%
j_{cLR}(\alpha _{r,l})$ shows up in the critical current $I_{c}$ observed in
experiment. Generally speaking, orientations of the $M$ vectors in F' and F
films are not necessarily the same ($\alpha _{r,l}\neq 0$). If there are
domains in the F' layers, the total critical current $I_{c,Av}$ is
determined by averaging the current density $j_{c}$ over the width of the
junction as the magnitude and sign of the average $\langle \sin \alpha
_{r}\sin \alpha _{l}\rangle $ depends on the number of domains, orientations
of the magnetization $M$ in these domains etc. This issue requires a further
theoretical and experimental studies.

\section{Appendix}

The boundary conditions for the matrices $\hat{f}_{0,3}$ follow directly
from Eq. (\ref{BC}) and have the form

\begin{equation}
\partial \hat{f}_{3}/\partial x|_{x=\pm L}=\pm \gamma _{B}\hat{f}%
_{S}|_{x=\pm L},\text{ \ }\partial \hat{f}_{0}/\partial x|_{x=\pm L}=0.
\label{A1}
\end{equation}

Substituting Eqs. (\ref{f0h},\ref{f3h}) into Eqs.(\ref{A1}), we obtain Eqs. (%
\ref{Sh}). The coefficients $C_{h\pm }^{(1,2)}$ are found from the matching
conditions of the functions $\hat{f}_{0,3}(x)$ and their derivatives at $%
x=L_{H}.$ They equal

\begin{equation}
C_{h\pm }^{(1)}=S_{h\pm }^{(1)}\frac{R_{\pm }(sc/cs)}{R_{\pm }(cc/ss)};\text{
}C_{h\pm }^{(2)}=S_{h\pm }^{(2)}\frac{R_{\pm }(ss/cc)}{R_{\pm }(cs/sc)}.
\label{A2}
\end{equation}

Eqs. (\ref{f3h}, \ref{Sh}, \ref{A2}) determine the form of the function $%
\hat{f}_{3}(x)$ in Eq. (\ref{EqF1}). The solution for Eq. (\ref{EqF1}) is
given by the formula

\begin{equation}
\hat{F}_{1}(x)=\hat{F}_{1un}(x)+\hat{F}_{1nun}(x),  \label{A3}
\end{equation}%
where $\hat{F}_{1un}(x)$, $\hat{F}_{1nun}(x)$ are solutions of a homogeneous
Eq.(\ref{EqF1}) (without the R.H.S.) and $\hat{F}_{1nun}(x)$ is a particular
solution of this equation. These functions are

\begin{eqnarray}
&&\hat{F}_{1un}(x) =\hat{\tau}_{2}[a^{(2)}\cosh (\kappa _{\omega }\tilde{x}%
)+b^{(2)}\sinh (\kappa _{\omega }\tilde{x})]  \nonumber \\
&&+\hat{\tau}_{1}[a^{(1)}\cosh (\kappa _{\omega }\tilde{x})+b^{(1)}\sinh
(\kappa _{\omega }\tilde{x})]\}  \label{A4}
\end{eqnarray}

\begin{eqnarray}
&&\hat{F}_{1nun}(x)=\sin \alpha \sum_{j=1,2}i(-1)^{j}\{\hat{\tau}%
_{2}S_{hj}^{(2)}[\frac{R_{j}(ss/cc)}{R_{j}(cs/sc)}\cosh (\kappa _{hj}\tilde{x%
})  \nonumber \\
&&+\sinh (\kappa _{hj}\tilde{x})]+\hat{\tau}_{1}[\frac{R_{j}(sc/cs)}{%
R_{j}(cc/ss)}\cosh (\kappa _{hj}\tilde{x})+\sinh (\kappa _{hj}\tilde{x})]\}
\nonumber \\
&&  \label{A5}
\end{eqnarray}

Matching Eqs.(\ref{A3}-\ref{A5}), we find finally the matrices $\hat{a}_{H}$
and $\hat{b}_{H}$, Eqs. (\ref{aH}-\ref{bH}).

\section{Acknowledgements}

We thank SFB 491 for financial support.

\end{document}